\documentstyle[manuscript,aps,version2]{revtex}
\begin{document}
\draft

\title
\bf  

Price Variations in  a Stock Market  with 
Many Agents 
\endtitle

\author{P. Bak$^1$, M. Paczuski$^1$, and M. Shubik$^2$}
\instit
$^1$Department of Physics, Brookhaven National Laboratory,
Upton, NY 11973\\
$^2$Cowles Foundation for Research in Economics, Yale University \\
 P.O. Box
208281 New Haven CT 06520 \\

\endinstit

\vskip -.5cm

\begin{abstract}

Large variations in stock prices happen with sufficient frequency to
raise doubts about existing models, which all fail to account for
non-Gaussian statistics. We construct simple models of a stock market,
and argue that the large variations may be due to a crowd effect, where
agents imitate each other's behavior. The variations over different
time scales can be related to each other in a systematic way, similar
to the Levy stable distribution proposed by Mandelbrot to describe
real market indices.  In the simplest, least realistic case, exact
results for the statistics of the variations are derived by mapping
onto a model of diffusing and annihilating particles, which has been
solved by quantum field theory methods.  When the agents imitate each
other and respond to recent market volatility, different scaling behavior is
obtained.  In this case the statistics of price variations is
consistent with empirical observations.  The interplay between
``rational'' traders whose behavior is derived from fundamental
analysis of the stock, including dividends, and ``noise traders'',
whose behavior is governed solely by studying the market dynamics, is
investigated.  When the relative number of rational traders is small,
``bubbles'' often occur, where the market price moves outside the
range justified by fundamental market analysis. When the number of
rational traders is larger, the market price is generally locked
within the price range they define.

\end{abstract}

\section{INTRODUCTION}

The literature on competition among firms and the literature on stock
market prices has been for the most part treated separately.  Here we
adhere to this approach, although eventually a model which makes
explicit the feedbacks between the ``real'' or physical economy of
firms and production must be linked to the paper economy of finance.
Even the simplest of economic models tends to become enormously
complicated when attempts at realistic modelling are made.
Mathematical tractability as well as clarity can be lost in a welter
of detail.  The basic idea of our investigation is to select model
segments of the estimated minimal structure needed to make an adequate
description of the statistical properties of price variations in a
stock market.

 A motivation for this work is suggested by the early observations of
Mandelbrot (1963, 1966,1967) on the nature of stock prices and the
recent behavior of the derivatives markets where ``10 sigma events''
have been happening with sufficient frequency to raise doubts about
existing models, which all fail to account for non-Gaussian variations
in price.  Mandelbrot observed that price variations of many market
indices over different, but relatively short time intervals could be
described by a stable Levy distribution, rather than being Gaussian.
The Levy distribution has substantially more weight for large events
than the Gaussian distribution where large events (beyond $\sim
4\sigma$) are prohibitively unlikely.  

Mandelbrot's observation has since been augmented by fitting to a
``truncated Levy distribution'' where the scaling regime for price
fluctuations is actually finite rather than infinite.  Empirical
observations suggest that the succession of daily, weekly, and monthly
distributions progressively converge to a Gaussian (Akgiray and Booth,
1988).  The presence of an intermediate scaling regime, where the
price changes $\Delta p$ occur with a probability distribution
$P(\Delta p) \sim (\Delta p)^{-\alpha -1}$, can mathematically account
for slow convergence to a Gaussian distribution at long time scales.
For example, the data for the S \& P 500 index is reasonably fitted by
a truncated Levy distribution with $\alpha \simeq 1.4$ over a time
scale which ranges from a minute to a day, with convergence to a
Gaussian at approximately one month (Mantegna and Stanley, 1995).
Also, Arneodo et al (1996) observed a $1/f^2$ power spectrum at long
time scales consistent with Gaussian behavior; while at short time
scales truncated Levy behavior was observed.  They analyzed the
DEM-USD exchange rate from October 1991 - November 1994.  Although
this Levy-type behavior is by now well documented empirically, there is as yet
no mathematical model of a stock market which can explain the origin
of large price variations with ``fat tails'', at least over a fairly
broad range of time scales.  Here such a model is introduced.

We construct an extremely simple, but completely defined, economic
model of many agents trading stock.  These agents form a market that
exhibits large variations in prices resulting from differences in the
agent's behavior.  The agents are of two types: type (1) are
``noise'' traders whose current volatility may depend on recent
changes in the market and whose choice of price to buy or sell may imitate
choices of others; type (2) are ``rational'' agents each optimizing their
own utility functions.  There is only one type of stock and each agent
can own at most one share.  These simplifications are quite drastic
for many reasons.  First of all, in reality
the price changes of different stocks
are correlated to each other.  Thus one cannot treat each stock as an
independent market.  In addition, we are ignoring price variations due
to exogenous changes such as in interest rates, money supply, or wars
breaking out.  Finally, we have an extremely simple description of
the individual agent's behavior.

In spite of these gross simplifications, some of our toy models exhibit a
statistical pattern of price variations consistent with that observed
empirically.  This suggests that we may have captured a dynamical process
that is sufficiently robust to describe  price variations in real
markets. One version of the model exhibits
 power law fluctuations at small time scales
leading to a
``Hurst'' exponent $H \simeq 0.6$.  Eventually at long time
scales the fluctuations may converge to Gaussian with $H=1/2$.  The 
fat tails in the probability distribution for price variations in
our model are a collective effect resulting from many different
interacting agents.  Our results suggest that large fluctuations in
price may be endogenous to the dynamics of stock markets.

\subsection{Review of Previous Literature}

Zipf (1949) wrote a provocative book in which he observed regularity
in the distribution of most frequently used words in English and
several other languages.  He also suggested that the size of an item
of rank $I$, by some measure was given by $I^{-\alpha}$ where $\alpha$
is an empirically determined exponent often near 1.  Zipf applied his
data analysis to diverse areas of human activity including the sizes
of cities, corporations, income of individuals, etc.  More recently
Mandelbrot (1963, 1966, 1967) in several articles noted that the
distribution of certain financial market prices appeared to be best
represented by a Pareto-Levy distribution.  In the tail of the
distribution the probability of price change of any size was present
as a power law.  A related observation was that the temporal
correlations in price fluctuations were characterized by a Hurst
exponent $H \simeq 0.6$ larger than that for a random walk ($H=1/2$).
Recently, the truncated Levy distribution has been used to describe
various economic indices, as described above.  These observations
have thus far confounded all attempts at a general explanation in
terms of a model economy or market.

Ijiri and Simon (1977) in a work covering articles written from 1955
to 1975 considered the size distribution of firms where the driving
mechanism is constant returns at any size, as suggested by Gibrat.
More recently Arthur (1989, 1990) looking at production has considered
models of market share based on what might be called ``network
increasing returns to scale.''  He considered the probability of
future market share to be dependent on the relative sizes of current
market share and offered a Polya process for updating the change in
probabilities of future market share.  Positive feedback leads to
``history dependence'' where the dynamics gives many 
possible outcomes for market share rather than a unique noncooperative
equilibrium solution. In yet another different but related approach
Bak and Paczuski (1995) discuss the role of power laws in many
sciences, including economics, noting the tendency of large dynamical
systems to organize themselves into a critical state with avalanches
or punctuation of all sizes.  Bak, Chen, Scheinkman and Woodford
(1993) have provided an elementary production and inventory model
which exhibits self-organized criticality (Bak, Tang, 
and Wiesenfeld, 1987,1988).
Based on an analogy
with traffic flow, Paczuski and Nagel (1996) have speculated that
a critical state with fluctuations of all sizes may be the most efficient
state that can be achieved in an economy.

It is with this body of literature as a basis that we attempt to set
up an elementary process model of interacting agents forming a stock market.
Ideally it would be desirable to construct a completely closed model
with the feedbacks between the producers and consumers fully
specified, but we suggest below that, as a first approximation, the
behavior of stock prices can be examined without going fully to this
level of complexity.

\subsection{Outline}

In Section II, the model is defined together with a description of the
agents and their strategies.  Section III notes the false dichotomy
between game theoretic and behavioral solutions which is often made.
We stress that there is no simple comfortable ``right way'' to
construct an expectations updating algorithm.  The standard
noncooperative equilibrium analysis finesses this problem of dynamics.

In Section IV, we present the results and an analysis of a series of
simulations of stock market trading with ``noise traders'' who may
imitate others as well as ``rational'' optimizing agents who make up
their own mind.  When the noise traders do not imitate, but act
in a purely random manner, we find that
the model corresponds to the universality class of the reaction
diffusion process $A +B \rightarrow 0$.   Barkema, Howard, and Cardy (1996)
have shown
analytically that scaling exists for the temporal variations
in the position of the A-B interface, which corresponds to
the price in our model, with a nontrivial Hurst exponent $H=1/4$
that is subdiffusive.  When the noise agents can imitate others, we
escape this universality class and find different behavior that is
more characteristic  of real markets.  In this
case, we obtain a broad distribution of price changes with an
effective Hurst exponent $H\simeq 0.6$ at short times, consistent with
a Levy or truncated Levy distribution, converging to a Gaussian with
Hurst exponent $H=0.5$. A particular interesting case is one where
the volatility of the market price is fed back to the agents.  An
apparent scaling regime exists with an effective Hurst exponent $H\simeq 0.65$.
The distribution of price changes is far from Gaussian, with a fat tail
of price variations far exceeding the Gaussian values,
very similar to data for real stock markets. 
  We also address issues as to whether the
rational traders can lock the price  in the case where noise traders
and fundamental analysts coexist, or if below a certain
concentration a ``depinning'' transition occurs with price bubbles
away from the rational expectations range provided by the optimizing
agents. 

Section V is devoted to a discussion of many of the open questions and
several promising approaches to the study of stock market price variations.
Section VI contains concluding remarks of our approach and potential
extensions.
 
\section{DEFINITION OF THE MODEL}

Models in finance and in monopolistic or oligopolistic competition
tend to be open models which are implicitly embedded in a large
outside economy (see for example Sharpe (1985)).  They are presented
without any particular need for either feedback from the outside or
conservation conditions on money to be satisfied.  In finance the
inputs from the firms behind the stock market are represented by
lottery tickets and the only commodity traded is money.  The
justification for this approach is that this simplification is
sufficient to capture enough detail to develop an adequate theory of
investment behavior.  This ignores the role of fundamental
securities analysis.  However,
 we follow this approach in order to examine for the
possibility that even at this level of simplicity some insight can be
gleaned concerning the nature of the distribution of price changes.

Prior to commencing any mathematical analysis, a justification for the
modelling choices made is provided.  In an actual economy there are
many different locations or sources of uncertainty.  In our models we
attempt to isolate several of these sources and to examine their
influence on fluctuations in stock prices.  Key to our approach is to
consider the dynamics of mass markets with diverse agents who may
imitate.
	
Heuristically we offer the following sketch of the extremely simple
economy we model.

\subsection{The Firm and Stock}

In our model, the presence of a single firm is only implicit in a
 lottery ticket or shares of the firm.  The ownership of a share gives
 the individual a dividend at time intervals $\Delta t$.  The size of
 the dividend is a random variable, for instance,
 determined   by a Bernouilli process
 which pays $A$ with probability $\rho$ and $B$ with probability
 $(1-\rho)$.  For simplicity we consider that each individual may buy
 or sell one share.  The share is traded {\it ex dividend}.

The dividend distribution affects the dynamics we are interested in
only through the utility function to be defined below.  However,
if we wish to do the bookkeeping for the profits and losses of the
traders, we assume that at each instant every agent who owns stock
is paid the average dividend. This regularization does not affect
our evaluation of the agents' performance in the long run where dividend
is paid many times. Of course, the liquidity of the agents is affected
by the random variations of dividend payments, but we assume that this does
not affect the agents' behavior.

\subsection{The Agents}

In the models below we consider two types of agents.  Both types are
gross simplifications but they provide enough structure to carry out a
reasonably complex investigation.  The first type may be characterized
as  noise  traders who learn essentially nothing about
equilibrium economics.  Each of these traders starts with an estimate
or ``guesstimate'' of a price at which he is willing to sell or a price
at which he is willing to buy a unit of stock.  At each time step,
a trader is chosen and will update his bid or offer by one unit,
approaching the current price with probability $(1+D)/2$ and moving
away from the current price with probability $(1-D)/2$.
Eventually an overlap occurs between a chosen buyer (seller) and
any other seller (buyer) and a transaction takes place.

The current  seller (buyer)  now becomes
a buyer (seller) at a new price.  Equivalently, one may think
of this as a new buyer and seller entering the market,
with the old ones that traded removed.  The new price is either chosen randomly
within a finite interval, or the new buyer (seller) picks another
buyer (seller) at random and copies the same price to buy (sell).
In general this type of
``urn'' or copying process is responsible for
 positive feedback and path dependence as noted
by Arthur (1990).  The urn process 
 also can create a peaked structure of organized agents
 leading to avalanches of all sizes
as found in a different context by Manrubia and Paczuski (1996).
Here we find that the positive feedback inherent in the urn mechanism
we are proposing gives rise to anomalously large fluctuations that
grow in time faster than a random walk over a broad range of time scales,
before eventually converging to the Gaussian result.

 Each type 1 individual is a simplistic trader who might lose money if
he churns too much at poor prices; but if the dividend rate is
sufficient, relative to the price he pays he could make money.  The
presence of dividends converts the process into a nonzero sum game and
all could profit if they did not overpay for a stock.  We assume that
there are $N-K$ traders of this type and $K$ traders of the next type.
There are $N/2$ shares available; hence half of the population are
potential buyers and half potential sellers.  The initial endowments
of an individual $j$ of type 1 are $(1, M(j, 0))$ if he owns stock,
and $(0, M(j,0))$ if he owns no stock.  Each individual who owns a
share is a potential seller and each individual who does not own a
share is a potential buyer.  It is not allowed to own more than one
share.  As we are not postulating an explicitly optimizing behavior
for these agents we do not need to  specify the (in general,
unobservable) utility function for each agent. Although if we wish to
``keep score'' we may do so.

\subsubsection{An Optimizing Agent with a Utility Function}

We assume that there is another type of trader who maximizes his short
run or period by period utility function which might be of the form
\begin{equation}
U=\nu \min [A,B] + (1-\nu)(\rho A + (1-\rho)B) \qquad .
\end{equation}
The interpretation of this equation is that each individual has an
overall risk averse utility function which is a mix of the extreme of
risk aversion $(\nu =1)$ and risk neutral $(\nu =0)$.  Indexing
$\nu$ as $\nu(j)$ for the $j$'th individual, we may give each
individual a different risk profile.  The traders buy or sell one unit
of stock or do not trade.  They invest any extra cash in a savings
account which pays a fixed rate of interest $i$ per period.

\subsubsection{Initial Conditions}

We assume that there is a fixed given rate of interest on
deposits = $i$. The $n (n = |N| )$ traders each start with a different price
expectation distributed in a rectangular manner over the interval $[p_-,
p^+]$. $(N-K)/2$ traders of the first type own a single share and $K/2$ of the
second type own a single share.

\subsubsection{Strategy}

In this highly simplified model no bank or other loan market is
assumed to exist.  We assume implicitly that all agents have more than
enough money to buy a share and that any money not spent in the market
is swept into an interest bearing account.  In a more complex and more
realistic model it is natural to introduce a commercial and central
banking system
designed to accept deposits or lend at a fixed rate of interest
$i$. This permits the system to create money in a model with an
outside bank. 

 The first type of agent has a decision rule which can
hardly be called a conscious strategy, but may have a tendency 
to follow the decisions of others in the market.  Thus when a buyer
who is an imitator
purchases a share and now becomes a seller, his initial offering price
is chosen by picking the price of a random seller.

 The second type of agent buys or
sells stock if he thinks its dividend return is high enough above or
below the interest rate.  This can be expressed as:

\begin{eqnarray}
{\rm Buy \ \  if}\qquad  & {U\over p_s} > i + \Delta_1 \quad ; \\
{\rm Sell \ \  if} \qquad  & {U\over p_b} < i + \Delta_2 \quad .
\end{eqnarray}
The parameters $\Delta$ reflect decision "stickiness". The distribution
of utility functions for the agents may be derived from a distribution
of $(\nu)$ and $(\rho)$ for the agents.  This leads to a distribution
of buying and selling prices for the heterogeneous agents. An agent's
 strategy
is to name two prices $p_s$ at which (or better) he will sell and $p_b$ at
which level (or cheaper) he will buy. These strategies can be regarded
as having been derived from an optimization on the one period utility
functions or otherwise considered merely as another form of
stock market behavior.

\section{GAME THEORETIC OR BEHAVIORAL SOLUTION?}

A simple false dichotomy between game theoretic multiperson
optimization approaches and behavioral solutions can be made.
This coincides with forgetting
 that the noncooperative equilibrium solution itself
is a behavioral solution.  When the individual
optimization is considered, it is contingent on expectations of
market price.  A specification of how expectations are handled cannot
be avoided.  Conditioning the individual's optimization on the
proposition that all individuals must have mutually consistent
expectations implies the acceptance of a weak form of local
optimization defined only at a fixed point in the system.

If one considers only equilibrium with consistent expectations
it is possible to engage in a complete finesse of all dynamics.  The
magic invocation of rational expectations is nothing more than another
way of stating that a noncooperative equilibrium exists.  It sidesteps
the problem of the formation of expectations by the argument that
given the appropriate expectations concerning prices the expectations
will be self validated by optimizing behavior based on their
acceptance.  Nothing is said about convergence to equilibrium if the
system is not already there.  No learning or other behavioral theory
is supplied to indicate how expectations are formed out of
equilibrium.

In essence, a game theoretic solution to an n-person game in extensive
form is nothing more than a set of n strategies which complete the
description of the motion of the system.  Any logically consistent set
will do, thus any well-defined behavioral model will suffice.  The
justification or rationalization as to why one wishes to consider one
solution better than another poses substantive questions about what we
wish to consider as rational economic behavior.

\subsection{Expectations}

As a game is played forward sequentially one cannot avoid
dealing with the formulation of expectations.  If one's only concern
were with equilibrium we can invoke a consistency condition requiring
that if an equilibrium exists all individuals' expectations must be
consistent so that no one is motivated to change strategy.  This may
be interpreted as the rational expectations assumption, or alternately
as postulating a noncooperative equilibrium.  However neither
assumption provides information on the nature of dynamics.  Even given
this condition, there is no guarantee that an equilibrium will be
unique.  In the approach here we consider both the game theoretic
solution of noncooperative equilibrium for the dividend optimizers and
a ``mass particle behavioral approach'' of the others.

\subsection{Information Conditions and Expertise}
 
A model of this variety may be highly sensitive to information
conditions.  The size of strategy sets increases astronomically with
knowledge of contingencies.  By letting an individual obtain
information about some of the randomizations he may be able to foresee
the future dividend of the firm. If information differentials are
introduced then the number of different types of players is increased
as they are now differentiated by information.

\subsection{Loans and Bankruptcy}

If it were possible for the individuals to borrow each period then we
would have to take into account the possibility that they could go
bankrupt each period after the first.  In these highly simplified
models we omit this feature which could increase market instability.
The question of bankruptcy is currently being studied by
Geanakoplos, Karatzas, Shubik, and Sudderth.  Even in a simple
fully defined process model with strategic choice the conditions
required to keep track of money and credit are somewhat elaborate and
tedious.  At first glance if there were a rule adding a heavy penalty
to the utility function it might appear that, for a stiff enough
penalty, all agents would strategically avoid default, but with
uncertainty present, as long as the penalty is finite, it is easy to
construct situations where there is a nonzero possibility that
individuals will borrow and default occurs.  When this happens rules
are needed to describe how the debt is extinguished.  Fortunately in
these simple models these difficulties are avoided by having no
borrowing and giving all plenty of money.

\subsection{Equilibrium and Behavioral Finance}

The noncooperative equilibrium and rational expectations studies are
devoted to the study of equilibrium positions.  By implicitly
requiring that expected prices and realized prices are always
consistent they throw away learning, error and all considerations of
dynamics. Even if one's only interest were to examine equilibrium
states of the system, a satisfactory economic model calls for a full
specification of a process model.  But with such a specification the
stage is set to consider adjustments from disequilibrium as well as
timeless equilibrium.  When markets are incomplete and there is
exogenous uncertainty present, at best we can only seek equilibrium
conditions in the aggregate, as the trajectory of any individual will
be random even if the assembly of agents shows some regular behavior.
The thrust of the work here is to select a set of models complex
enough to examine dynamic behavior of many
competing agents  yet, in some instances simple
enough that we can examine the noncooperative equilibria and contrast
them with the behavioral models.
 
\subsection{A Few Crude Facts}

In order to sweeten our intuition for this type of model, a few
statistics selected from the New York Stock Exchange (NYSE) 1994 Fact
Book supply some orders of magnitude.  In 1994
approximately 51,000,000 individuals in the United States  owned
stock, and around 2,600 issues were traded on the NYSE. The firms
had 140,000,000,000 shares valued at \$4.45 trillion, giving an
average value of around \$31 per share and an average issue of around
53,000,000 shares per firm.  There were around 10,000 institutional
investors or intermediaries and around 120 initial public
offerings raising \$22,000,000,000 or approximately 1/2\% of the total
market value.  The velocity, or turnover ratio was 54\%.

\subsection{A Trading Volume Puzzle}

We observe that in the actual New York market the volume of trade has
been of the order of more than 50\% yet, as is noted in the example in
Section IV.A, if we postulate that there is a single type of trader
with indefinite life, the noncooperative or rational expectations
equilibrium will involve all individuals holding an optimal portfolio
of stock and never trading.  What are the factors that account for
stock market trading and how large are they?  We consider five
factors.  They are: 1. overlapping generations (OLG); 2. life cycle
considerations; 3.  stationary risk trading; 4. the influence of
exogenous events and 5. heterogeneous investment activity with
nonconsistent expectations.  If we consider that the period of time
for which an individual owns a portfolio of stock
 is somewhere between 20-50 years,
then the pure OLG contribution to stock trading for a constant population
is somewhere between 2-5\% on the conservative assumption that stock
will be sold rather than kept from generation to generation.  
Even a
casual glance at life cycle changes in expenditure such as selling
stock to put  children though college or selling off stock to
compensate for loss of earnings in retirement probably does not
account for more than 10-20\%.  

Karatzas, Shubik, and Sudderth (1995)
and Duffie, Geanakoplos, Mas-Colell, and McLennan (1994) utilizing
somewhat different models were able to show the existence of
stationary strategies for an economy modelled as a set of parallel
dynamic programs.  Thus there is a stationary wealth distribution for
all agent which maps onto itself.  The same agent in such a stationary
state could be poor and risk averse at one point in time and rich and
possibly risk neutral at another point in time.  It is possible that
agents changing wealth status in a stationary state might wish to
exchange portfolios. Even if we could establish this as
feasible, an open question remains as to what proportion of
share trading could be attributed to this type of trade. 
Thus there is a basic question concerning how much of stock trading
can be explained by equilibrium analysis and how much requires
explanation from the dynamics of nonstationary trade.

Economic systems are subject to exogenous shock such as wars breaking
out, presidents being shot, natural disasters occurring, or innovations
creating new markets and wiping out old ones.  All of these
incidents, in one form or another, could impact on different
individuals with different levels of intensity and lead to
considerable arbitrage before the system settles down (if it ever
does).  It  is not known how much trade is caused by these phenomena.  The
fourth and fifth points are connected.  Even if we believe fully in
Bayesian updating, no matter how complicated and in rational
expectations or mutually consistent expectations with noncooperative
equilibria, all of this apparatus does not tell us where our {\it a priori}
expectations of when the next earthquake will occur or who will be the
presidential candidate in eight years time and how these events will
impact the stock market.  The way the phrase ``rational behavior'' is
used in much of economic theorizing is as a euphemism for one person
optimization in a context given scenario.  But much of economic
activity is devoted to getting the context right.  Thus the key
question not asked in the rational expectations or noncooperative
equilibrium analysis is how do individuals obtain a cognitive map of
their problem, form expectations, and estimate consequences.  The
second question, answered by those who accept rational expectations
analysis is that, given the context all individuals act as if they are
maximizing their expected payoff in a one person optimization problem.
The maximization assumption is essentially normative and although it
might hold in circumstances where individuals are experts and
dedicated money maximizers it does not necessarily provide a good
empirical view of actual behavior.  If there is a substantial segment
of the market following some other rule its presence could
influence the environment in which the ``rational optimizers'' act.  Our
investigations in Section IV consider this possibility.

\section{Simulations of the Market}

Before discussing the various models which have been studied numerically,
we first
define the mechanics of the market in which the agents are
playing. There is one type of stock, and each agent can own only one
share.  Therefore, each agent is either a potential buyer (if he
does not own a share) or a potential seller (if he owns a share).
For simplicity and without loss of generality, we assume that
there are $N$ agents and $N/2$ shares. Each share owner advertises a
price, $p_s(j)$ at which he is willing to sell his share, and each
potential buyer advertises a price $p_b(j)$ that he is willing to pay
for a share. The prices may assume any value in the interval
$0<p<p_{max}$. The various types of agents differ in the way that
these prices are chosen.  

At each update step a single agent is
chosen randomly to look into the market. For instance, imagine that he
 turns on his computer and observes all advertised prices in the
market. The
market is a list of advertised buying and selling prices for all
agents.  If the agent owns a share, and observes that there are one or more
buyers who are willing to pay more than his advertised price, he sells
to the buyer offering the highest price. That price defines the market
price, $p(t)$, of the stock at that particular instant when the
transaction occurred. 
The seller  (buyer) in the transaction now becomes a potential buyer
(seller) and chooses
a new price $p_b$ ($p_s$) at which he is willing to buy (sell) a share.
If no transaction occurred,
the agent may update his advertised price according to his particular
rule, which may or may not be related to his observations of the
market.

If the randomly chosen agent did not own
a share, and there are sellers willing to sell at or below
the price which he is willing to pay, he buys from the agent charging 
the lowest price, and now becomes a seller
who advertises a selling price $ p_s(i)$. If no transaction occurred, the
agent  may change his bidding price. This process is repeated {\it ad 
infinitum}. For simplicity, in the figures presented here,
a time unit is chosen as the average number
of updates for an agent.    Thus, in $t$ time units each 
agent has been active and looked into the market on average $t$ times.
Since the updating of the agents is (random) sequential there is no need for an
intermediate agent clearing the market, as would be the case for a situation
with many simultaneous bids.

\subsection{A Market with Fundamental Value Buyers Only.}

Let us first consider the situation where all agents are ``rational'',
choosing their prices according to the utility function defined in
Section II.B.1.  Each agent $j$ has a different stickiness $\Delta(j)$
and risk aversion $\nu(j)$.  These variables are chosen from some
arbitrary distribution which is bounded between zero and one.  One
sees from Eqs. 1-3 that this translates into different buying and
selling prices, $p_b(j)$ and $p_s(j)$, for the agents with a
distribution which is determined by the parameters in Eqs. 1-3.
The buying and selling price for each agent are ``quenched'' random variables;
i. e. they remain the same throughout the simulation.
The details of the distributions of $p_b(j)$ and $p_s(j)$ are not
important.

We assume that the
distribution of buying prices for the agents
 is uniform in an interval which partially overlaps
with a higher interval of uniformly distributed selling
prices. After some time, the market comes
to rest in a state where all $N/2$ agents who own stock  have
a selling price
$p_s$ above a certain price $p^*$, and all the $N/2$ agents who do not
own any shares have a buying price
$p_b$ less than $p_*$, where $p_* < p^*$. Because
of the stickiness $\Delta$ the position of the gap between buyers and
sellers depends on the initial conditions.  In this state, none of the
agents have any reason to take further action.  Buyers and sellers
``phase separate'' into nonoverlapping regions of price with a dead zone
in between.  Thus, with only
one stock for sale, rational agents will not trade in equilibrium
but hold their portfolios indefinitely.  In all the numerical simulations
to follow, we shall assume that the rational agents had already reached
such an equilibrium state with no trade at the beginning of the simulation.

\subsection{A Market with Simple ``Noise Traders'' Only.}

At the other extreme, let us consider a market where all the agents
trade according to their observations of the state of the market
without concern for the fundamental values. In principle, their
behavior can be defined in a number of different ways. For simplicity,
we may initiate the simulation in a state where the $N/2$ agents who
own stocks are willing to sell at prices which are uniformly
distributed in the interval $p_{max}/2 < p_s < p_{max}$, and the bids
of the $N/2$ agents who do not own stock are uniformly distributed in
the interval $0 < p_b < p_{max}/2$.

There are many different ways to define the behavior of the noise
traders.  Perhaps the simplest behavioral model is when each agent's
price fluctuates randomly, independent of the other selling or buying
prices in the market.  The agents interact only by buying and selling
when there is an overlap in prices.  We consider this simple model
first, before investigating more complex behavioral models with a
higher degree of interaction between agents.

\subsubsection{Independent Noise Traders}

At each update step, the price $p_s(j)$, if the chosen
agent $j$ is a seller, or $p_b(j)$, if agent $j$ is a buyer,
changes 
randomly by one unit with equal probability 
in either the downward or the upward 
direction.  Most of the time this will not induce any trade, but 
occasionally  the chosen agent  will find himself at a price level
 where other buyers or sellers 
become interested, and a sale will take place at a price $p(t)$. After the 
sale, the new buyer will choose a new bidding price for possible future 
trade randomly between 0 and $p(t)$; the new seller will choose an 
asking price randomly between $p(t)$ and $p_{max}$. How does the price $p(t)$ 
vary with time?

It turns out that a very similar process has been studied extensively
by physicists, and a mathematical solution exists for the fluctuations
of $p(t)$ which are scale free.  The Independent Noise Traders model
is equivalent to a reaction-diffusion model where two types of
particles, A particle and B particles, are injected at different ends
of a tube.  These particles each diffuse randomly until a particle
bumps into a particle of the opposite type, whereupon they annihilate
each other.  When this reaction occurs new A and B particles are
injected at opposite ends of the tube (Figure 1). The positions of the
two types of particles correspond to the prices $p_b$ and $p_s$
respectively. The position where the annihilation event takes place
corresponds to the market price $p(t)$. The reaction-diffusion
process is called ``A+B
$\rightarrow$ 0'', and has been extensively studied.

Barkema, Howard, and Cardy (1996) have
 shown mathematically that the variation $\Delta p(t)$ 
of the price after time $t$ 
scales as 
\begin{equation}
\Delta p(t) \sim t^{1/4} (\ln (t/t_0))^{1/2} \quad ,
\end{equation}
where the parameter $t_0$ could depend on the number of particles $N$ and
the size of the price range or ``tube'', $p_{max}$. Thus, at very long
time scales the price variations are scaling with
a Hurst exponent $H=1/4$.  This value is much  less than
the exponent for a random walk $H=1/2$, where the distribution after a long
time is a Gaussian with a width scaling as $t^{1/2}$. For shorter
time scales the variations are larger because of the logarithmic factor
in Eq. (4). 
Of course for
the longest time scales the variations cease to increase because of the global
confinement of the price range.
We stress that the situation considered here is pretty academic
in the context of economic theory. Nevertheless,
it demonstrates rigorously that anomalous scaling behavior, like the one
observed for real markets, can arise as a consequence of the interactions
between very many agents in a simple market model. This type of scaling
cannot occur in systems with a few degrees of freedom, or agents in our case.
The formalism used to derive Eq. (4) makes use of Quantum Field Theory,
and there is no simple intuitive argument for the anomalous (non-Gaussian)
form.

\paragraph{Numerical Simulation Results}

The behavior of the Independent Noise Traders model
 is illustrated in Figure (2 a,b,c,d). The system includes
$N=500$ agents operating within a price range of $p_{max} = 500$. Figure 2a 
shows the variation of the price vs time. 
Figure 2b shows the histogram for
the distribution of agents' prices 
at some instant in time. The 250 agents to the left of the gap at $p=247$ are 
currently potential
buyers, the 250 agents to the right of the gap are sellers. Note that the 
number of potential buyers or sellers is relatively small near the gap 
which defines the current market price. Agents diffusing into this regime will 
trade, or ``annihilate'', and be shifted to other values on the other
side of the gap.

The price variations can be conveniently represented by a 
Hurst plot (Feder, 1989). The variation over a time interval
$t$ is 
characterized by the range $R(t)$, which is the maximum 
variation over nonoverlapping time intervals of length $t$, averaged 
over the entire time record of the simulation. For Gaussian processes $R(t)$ 
increases as $t^{1/2}$. On a log-log plot the curve for
a Gaussian process would be a straight line 
with the slope equal to 1/2. Figure 2c shows $\log R$ vs $\log t$ for the 
simulation.  From the theoretical result for the
equivalent A + B $\rightarrow$ 0 process,
we predict $\ln R = 1/4 \ln t  +  \ln\big(\ln (t/t_0)\big)$. Indeed, 
for large $t$ the slope approaches 1/4 asymptotically, and for smaller $t$ 
the apparent 
slope is larger because of the logarithmic corrections. The logarithmic factor
represents the large noise or glitches
at shorter time scales which overlay the slowly
varying long
time power law behavior, as seen in Figure 2a. Thus, for 
small $t$ the effective exponent for the fluctuations in Figure
2c is much larger 
than 1/4, and remarkably even larger than the random walk value of 1/2.  The 
logarithmic factor gives rise to slow convergence to the asymptotic value
1/4. Thus, if one had access only to short time fluctuations one might 
erroneously conclude that the exponent is much greater than 1/4. 
In Figure 2d, we have plotted both $\big(R^2(t)/t^{1/2}\big)$ and
$\big(R^2(t)/t^{1/2}(\ln t/t_0)\big)$, with $t_0=1$, and find
results consistent with Eq. (4).

The price variations in the independent noise traders model
 asymptotically have a power-law form discussed by Mandelbrot,
 although the exponent is smaller rather than larger than
 1/2. However, at short time scales the model exhibits larger than
 Gaussian fluctuations as a result of the logarithmic term.  These are
 both rigorous results.  

One might naively have suspected that the
 model would exhibit pure random walk behavior for the price
 variations, since it is based on the diffusive behavior of the
 individual agents.  Nothing could be further from the truth.
 Although the price variations of this model are not consistent with
 observations of reality, we find that the price variations exhibit
 scaling with a logarithmic factor responsible for slow convergence to
 the asymptotic result.  There is excellent agreement between our
 numerical results and the previous analytic work on the
 reaction-diffusion system.  We now consider what effects are able to
 alter the universality class of the model and take it outside the
 regime of the A + B $\rightarrow$ 0 system.

\subsubsection{Drift Toward the Current Price}  

We now allow the agents to adjust their current price towards the
actual price of the last trade that occurred in the market.  The
agents become more realistic as time passes, but again, their behavior
is unrelated to any fundamental value of the stock. At each update
step the agent that is chosen moves one step towards the market price
with probability $(1+D)/2$, and one step away from the market price with
probability $(1-D)/2$. In the simulation, a drift coefficient $D=
0.05$ was chosen. Figures 3, a, b, c are equivalent with the
corresponding figures for the purely diffusive case of independent
noise traders. A finite drift does not change the general pattern,
including the slow convergence towards the exponent $H = 1/4$ for the
price variations.  

The slow price variations are related to the fact that the prices are
confined by fiat
 to a box of size $p_{max}$. These noise traders at least realize
that prices have to be within a certain range - otherwise they would
have no clue to choosing their prices between zero and infinity. After
a trade, the current buyer and seller would choose a new price
randomly. A more
interesting situation arises if the agents, when choosing their asking
prices and bids, mimic existing traders in the market. This type
of urn process can lead to large scale organization in dynamical systems
(Manrubia and Paczuski, 1996). 

\subsection{The Urn Model}

We will consider an ``urn'' process, where after each
transaction the new selling and buying prices for the trading agents
are chosen by randomly picking a buyer and seller and copying his
price.  This has the effect that the new price is chosen
with a probability proportional
to the number of traders who currently have that price.  However,
no global information on the part of buyers and sellers is needed when
choosing a new price. This opens up the possibility of mimicking crowd
behavior, where agents follow each other, again without paying any respect to
fundamental market values.

We modify the previous model of noise traders with drift where the
agents, in addition to copying each other's price choices upon
becoming a buyer or seller, exhibit a stochastic drift towards the
current market price. This allows for the preservation of organized
clusters of agents with similar prices. Once the market drifts away
from its ``true'' value the price is not subject to
any restoring force whatsoever. 

\subsubsection{Numerical Simulations of the Urn Model}

Figure 4a shows that the price variations in the urn model with
$N=500$ agents are much more dramatic than in the previous case, and
qualitatively look a lot more like the variations observed in a real
stock market.
Initially, the buying prices $p_b(j)$ were chosen randomly between
1900 and 2000, and the $p_s(j)$ between 2001 and 2100. The parameter
$p_{max}=4000$, although it could be made arbitrarily large.  The
collection of agents organize themselves into a well defined price
range (Figure 4b) which is not affected by the external parameter $p_{max}$ as
long as $p_{max}$ is sufficiently large.  This is in contrast to the
case of independent noise traders (see Figs. 2b, 3b) who fill up the
entire interval $[0, p_{max}]$.  At very long time scales, the center of
mass of this
self-organized distribution of market prices wanders up and down as a
random walk. The exponent for the variations is $H=1/2$ at long time scales.

Again, though, the convergence is very
slow. Although we presently have no analytical results to refer to,
we speculate that  the
slow convergence may be due to  a logarithmic factor as for
the previous model with $H=1/4$. For
relatively small $t$, the effective Hurst exponent is larger than 1/2. Indeed,
as referred to in the Introduction, 
it has recently been argued, based on real economic data, that in the
long run price variations are random walk.  The apparent Levy
distributions with $H>1/2$ are a transient phenomenon for
short time scales only, but there is slow convergence to
the Gaussian process.  Remarkably we observe
a very similar slow convergence to Gaussian behavior, with
superdiffusive fluctuations at small time scales, in  simulations
of the urn model.

\subsubsection{Volatility Feedback}

Very interesting behavior was observed in the case where the information
on the volatility of the market is fed back to the agents.
It is known that that various market indices exhibit
volatility clustering. At times 
where prices  have recently been volatile, this volatility could influence
the behavior of the noise traders, or the imitators. 
If the Dow-Jones index exhibits a 
large drop on a given day, it is likely that there will be a large 
variation the next day, although it could go either up or down. We 
attribute this to the reaction of the agents to the price variations. To 
mimic this effect, we simulated the urn model above
with the added feature that the diffusion constant and drift
for the agents' prices is proportional to the actual recent variations of 
market prices. 

In the simulation, if the price change during the last period of 100
time units, is $\Delta P$, an agent updating his price will increase
or decrease his price randomly by an amount $\Delta P$.  The probability
for increase is $(1-D)/2$ and decrease is $(1+D)/2$ as before.
Thus, instead of moving one unit up or down, the agent moves
$\Delta P$ units.   Since $\Delta P$ depends on recent changes in
the market price, this obviously is a mechanism for positive feedback
in the economy where large fluctuations create space for large fluctuations
in the future. 

 Figure 5 shows the price variations in the urn model with volatility
feedback. Indeed, the variations are more dramatic than in any of the
other cases studied.  As shown in Figure 5c, there is an apparent
plateau with an exponent $H = 0.65$ even for relatively long time
scales. 

In order to study the scaling behavior further, we plot the
distribution of price differences $dp$ at a fixed time interval $dt$
for various values of time intervals.  This was done for real markets
by Mandelbrot (1963, 1966, 1967), Mantegna and Stanley (1995), and
Arneodo et al (1996).  One simply samples the index, e.g. the S\& P 500, at
regular intervals (say $dt$ is one hour) and records the differences
between subsequent measurements. Figure 5d shows the price difference
distributions for time intervals $dt$ ranging from 200 to 6400 time
units in our simulation.  Of course, the larger the time interval, the
larger the variations.  However, when plotted in terms of properly
defined scaling variables, the picture becomes immensely
simplified. Figure 5e shows the same data plotted as a function of the
scaling variable variable $z=dp/(dt)^{0.65}$. In the Figure, the
vertical axis is scaled so that the entire distribution is
normalized to 1.  Within statistical
uncertainty, all the curves collapse onto a single curve. This data
collapse shows
that the distribution of price variations $P(dp,dt)$ exhibits scaling
behavior, 
\begin{equation}
P(dp,dt) \sim F \bigg(dp/(dt)^{0.65}\,\bigg) \quad , 
\end{equation}
i. e. it can be expressed in
terms of one scaling variable instead of two independent variables.
This was precisely the behavior observed by Mandelbrot, who suggested
that the scaling function $F$ is a stable Levy distribution.

Figure 5f shows
the scaling function as obtained for $dt=1600$. For comparison a
Gaussian with the same variance is also shown. Note the dramatic
difference. In particular, the distribution shows a fat tail
indicating a significant probability of observing fluctuations which
greatly exceeds the standard deviation. The data is insufficient
to determine whether or not the scaling function is precisely a Levy
function with power law decay. The value of the exponent in the
scaling function is very
similar to the one observed for real markets. The time intervals over
which there is scaling spans the interval from $dt=200$ to
$dt=6400$. The upper limit corresponds to a time interval during which
each trader on average performs a few trades. The position of the
center of mass of the
distribution of agents, Figure 5b, is not significantly shifted during
that interval.  At larger timescales, the distribution shifts, and the
price variations must become Gaussian. This simulation is our most
convincing argument that the scaling of price variations in real
markets indeed has its origin in the collective ``crowd'' behavior of
very many agents interacting with each other by  imitating
and watching the same
market data, irrespective of underlying fundamental values.

In the remaining numerical simulation studies we do
not include the volatility feedback effect in the noise traders,
but revert to the simple urn model.

\subsection{A Market with Fundamental Value Buyers \\ and Noise Traders
with Imitating Behavior}

It is natural to consider what happens in a market with both rational
optimizing agents and noise traders who imitate others.  In a real
market situation, we might expect both types of players, as well as
many other types.  In our computer laboratory, we can ask some
basic questions about the market.
For example, will the existence of rational players be sufficient to
discipline the noise traders so that they can get a free ride, or will
the rational agents be able to systematically exploit these traders?
Maybe the noise traders change the dynamics sufficiently, so that the
rational traders have to take their behavior into account.

In the initial setup, the noise traders' prices are distributed in an
interval around the median $p^*$, limited by $p_{min}$ and
$p_{max}$. Half of the rational traders own a share and their asking
prices are distributed randomly in an interval above $p^*$. Their
potential buying prices, which become relevant if they happen to sell
their share is also fixed throughout the simulation.  They are an
amount $\Delta$ lower, with $\Delta$ distributed randomly between 1
and $\Delta_{max}$. Similarly, the second half have buying prices in
an interval below $p^*$, and if they happen to buy, their potential
selling price will be an amount $\Delta$ higher than their buying
price.  The noise traders are not aware of who is a noise trader and
who is rational.  When they are selecting new prices after buying or
selling, they randomly choose the price of any other agent, so that
both the prices of rational and noise traders are copied.  In addition
to the actual transaction, this provides a coupling between the two
types of traders.

\subsubsection{Numerical Simulation Results}

Figure 6 shows the results for a simulation with a small fraction,
2\%, of rational traders.  All the prices of the rational traders are
arbitrarily confined to the interval between 1931 and 2068. This interval is
supposed to represent the interval spanned by the utility function in
Section II.B.  The highest price can be thought of as the price with
zero risk aversion, plus the stickiness $\Delta$.

 In the beginning of the simulation,
the prices of the two types of agents are similar, but eventually the noise
traders convince themselves that the stock is worth more than it is,
and their prices escape to higher values. At that point, all the
rational traders have sold their shares; they are
out of the market.  Eventually the market price returns
to the window of the rational players. This behavior, where the
crowd effect leads to unreasonably high prices is known as a ``bubble''.
At least in our model, we observe that
 this is a possible outcome in a market
with few rational traders. The price variations at long
time scales follow a power law
with exponent $H=1/2$ just as for the case with noise traders only in
the urn model.

Figures 7a,b shows the situation with 20\% rational traders. Now, the 
price range of all traders is confined within the range of the 
rational traders. Only very brief deviations occur, and common sense 
is rapidly restored. Figure 7b shows the skew distribution of prices 
during the brief excursion that occurred around 4000 on the horizontal
axis of Figure 7a.  At that point, all of the fundamental traders own stock
and most of the noise traders are undervaluing the stock.

\subsubsection{Performance Analysis}

De Long, Schleifer, Summers, and Waldmann (1990), in a stimulating
analytical paper, have considered an economy with optimizing agents
and noise agents.  In their analysis they observed that
under some circumstances the noise agents might outperform the rational
optimizers.  If the optimizers are risk averse, the presence of random
agents adding to the variance in price will make the optimizers more
cautious that they would be otherwise.

We have measured the performance of all the agents in the simulation,
extending over 1 million time units.  During that interval, each agent
on average was updated 1 million times.  The noise traders
have traded on average 4100 times, while the fundamental value traders only 80
times. Thus, their activity is only 2\% of that of the noise traders. Actually
5 of the 100 rational traders did not sell at all during the simulation.

There are two sources of profits, dividends and capital gains.	
The average capital gain for the fundamental value traders was 187,
compared with a loss of 35 units for the noise traders. The rational
traders made their capital gains simply by buying low and selling
high, whenever possible, and doing nothing in the meantime. Some of
the rational agents were lucky enough to make significantly more that the
average due to a strategically favorable choice of buying and selling prices
near the equilibrium around 2000. One agent made a profit of 1029
units. The noise traders share the losses in a much more democratic way, with
no significant variations from agent to agent because their prices
are changing all the time while the rational agents prices are quenched and
fixed throughout the simulation.

In order to estimate the dividends relative to the interest rate, we
assume that the highest selling price, $p=2150$, among all the rational agents
is the price for which the dividend, on average, balances the interest
rate. Thus, if you buy a share for $p=2150$ you make no profit from
dividends. If
you buy at a lower price $p$, you make a profit by keeping the difference
$2150-p$ in the bank. With an interest $i$ paid every $t$ time steps, the
profit becomes $(2150-p)i/t$ per time step. Assuming somewhat arbitrarily that
$i/t=1/1000000$, the average profit from dividends for the noise traders
is 72 units, the average profit from the rational traders was 83 units.
Thus, there appears to be no significant difference.
 This is due to the fact that
all the trading
action is rather close to the equilibrium value of 2000, so the average
profit from dividends
for all agents during the 1000000 time steps is 75, corresponding
to half the agents holding shares and making profits at any given time.

However, this statement is deceiving. The profits from dividends of
some rational agents were twice as large, obtained by simply holding
on to the stock throughout the simulation.  Others did less well by
offering too low prices and never entering the market.
  The time scale for the total simulation can
be thought of as the time scale for which the interest payment is
equal to the stock price, of order 10 years.  The time unit is
1/1000000 of that, of order 5 minutes.

\subsection{A market with Experts Introduced}

We may model certain aspects of expertise by considering that some of the
agents are more adept at ``picking winners'' than others.  A way in
which aspects of expertise can be modelled is to let some
individuals know the outcome of certain random variables
when others are not yet informed.  This, in essence, enlarges their
strategy sets.  If we did this here the advantage goes to the better
informed.   The strategy sets of the experts are larger than the
other optimizing agents as objective uncertainty about next
period's dividend is removed.  However, their uncertainty about the
behavior of the noise traders still remains.  Keynes' observation about
wanting to know what the average opinion of what the average opinion
will be still holds.  Dubey, Geanakoplos, and Shubik (1987) were
able to show analytically the gain to the better informed relative
to the less informed of the same type in a game with rational
traders only.  We do not simulate variations in expertise as this analysis
suggests that qualitatively, beyond observing that the more expert
of a specific type will do better than the less expert, no new
phenomena will appear.

\subsection{Survival of the Fittest?}

A simplistic view of competition would be that the optimizers and experts
would eventually drive out the less skilled.  However, the stock market
is not a zero sum game and, as in predator-prey species relationships,
it is feasible for both species of traders to survive, as seen in the
numerical simulations above.  Even in games
which are zero sum in money, there are after all individuals who buy lottery
tickets or play slot machines for their entire lives, while others
sell lottery tickets or own slot machines.

\subsection{Trading with Many Stock}

For simplicity we have limited our considerations thus far to trading
in one security.  A natural generalization of our modelling would
be to consider trade with many securities.  We note that a distinctively
new phenomenon appears in proceeding from one to two or more stocks.
The noncooperative equilibrium of the strategic market game for
the optimizers only may longer be unique even with trade in only
two different stocks, both of which pay out in ownership claims to
the same single consumer good.  Furthermore at an equilibrium point,
there may be stock trading taking place, unlike the case for a single
stock.  We defer the investigation of trade with many stocks to future
investigation.

\section{Many Models of Market Pricing}

The models presented in this section provide an opportunity only to
scratch the surface of a highly complex set of problems.  In this section
we briefly comment on other approaches and phenomena which merit
consideration.

\subsection{ Experts, Intermediaries, and Social Networks}

Another potential source to consider as contributing to the dynamics
of stock market prices may come through the use of intermediaries.
One of the little understood phenomena in social psychology is the
behavior of the crowd.  The classical work of Le Bon (1982) written
over a hundred years ago is still possibly the most perceptive writing
on the nature of crowds.  When a panic breaks out it is as though a
mass of individually independent particles were all simultaneously
polarized.

In the few statistics noted in Section III.E we observed that there are
around 10,000 financial intermediaries trading on the NYSE.
Institutions such as pension funds, mutual funds and insurance
companies hold well over 50\% of all equities (see NYSE Fact Book,
1994, p.83).  More and more the individual investor invests through an
intermediary who supplies credit, liquidity, matching, accounting,
record keeping, transactions costs savings, information and, in some
instances, expertise.

A natural extension of our models is to include an extra layer of
intermediaries. An extreme model has all 50,000,000 individual agents
as secondary sensors who spend their time trying to select the best of
the 10,000 intermediaries who spend their time trying to buy the best
portfolio of the 3,000 firms.  The variation in expertise lies
primarily in the 10,000 intermediaries.  Historically, many of the
intermediaries were partnerships or mutual firms, but recently many
have become corporations with stock traded on the major exchanges.
Banks and insurance companies appear, on the whole, to have been longer
lived than stock brokerage houses, where often success depended on one
expert individual.  As a first crude approximation one could have a
birth death process for a financial intermediary to be the same as the
manufacturing firm.  Left out of these models are several further
basic sources of fluctuations in the economy.  They are the role of
credit granting (in 1994 margin debt alone was around \$60 billion or
around 2.5\% of trade), and the feedback between the firms and the
stock market.

The network increasing returns to scale of Arthur (1990), the study
of imitation and social learning (Gale and Rosenthal, 1996), and
the work in contagion all suggest the importance of communication networks
in correlating human behavior.

\subsection{Process Models, Dynamics, and Heterogeneous Agents}

A key theme in our approach is the necessity to have a fully defined
process model in the study of stock market trading.  Much of the
misunderstanding about rational expectations dissolves when a
consistent process model of trade is considered.  Recently in the
study of economics, there has been a growing realization of the
importance of studying models with agents that have heterogeneous
characteristics (see for example Grandmont, 1993), expectations, and
strategies (Arthur, Holland, LeBaron, Palmer, and Tayler, 1996; De Long,
Shleifer, Summers, and Waldmann, 1990).  The assumption of rational
expectations or noncooperative equilibrium, by imposing prematurely
consistency conditions on expectations, removes the degrees of freedom
present in the actual dynamics, which could result in many different
trajectories.  The empirical question of how people form expectations
is finessed and replaced by a convenient mathematical simplification
that has to be rationalized as a ``good enough approximation''.

\subsection{Learning, Habit, and Optimization}

In the work of Arthur, Holland, LeBaron, Palmer and Tayler
 (1996), the agents
are highly adaptive, living in a soup of different conjectures and
rules of thumb which they test incessantly against each other.  Simon,
years ago, coined the term ``satisficing'' to indicate behavior where
habit takes over from active search.  In finance there are the ideas
of the market niche and financial boutiques.  In the study of
innovation and new products, since Schumpeter, it has been observed
that it is not uncommon for a previously successful industry to die
out because the lesson learned yesterday that led to success is not
unlearned sufficiently fast to permit adaptation.  There is a trade
off between the variability in the environment and the flexibility of
learning.  If one can find the appropriate niche, then the amount of
learning needed to survive may not be too high.  In the markets the
long term investors, bankruptcy experts, short sellers, day traders,
and bond traders all have different niches in both location and
timing.  Peters (1994) in his consideration of fractal market analysis
suggests different timing niches as critical to the trading structure.

\subsection{The Importance of Time and Size Scales}

In the study of physics, several different physical and time scales
are identified.  The laws governing interactions among the extremely
small and the extremely large may differ from each other and from that
which lies in between.  Observations and analysis
of emergent phenomena where the
large scale behavior can not be predicted from the details of the small
scale dynamics are the source of much of the contribution that physicists
might make to the study of complex systems in economics or other fields.

In the study of economic activity, there appears to be natural lower
bounds to the length of time in which a trade can take place.  This
represents a small scale cutoff in time.  An upper bound reflects the
time for which an individual is concerned with his or her trading.
This represents a large scale cutoff in time.  The lower bound is
supplied by the minimum amount of time required to decide on a trade,
transmit the decision, have it executed, and verify that it took
place; the upper bound is the individual's length of life.  

Much of microeconomics (and our examples studied in Section IV) is
concerned with open systems.  Convenient assumptions such as unlimited
availability of credit and no bankruptcy are made in order to avoid
having to consider conservation laws on money and credit or the
boundary conditions caused by bankruptcy.  Clearly, money is quantized
at the smallest scales, and the total supply sets a large scale
cutoff.  When the economy is considered as a whole these details
cannot be ignored.  In economics, the scope of the model may require
new considerations.  Feedbacks and restorative forces working at one
scale may have little, if any influence, at a different scale.

\subsection{Quantitative and Qualitative Aspects of Uncertainty}

In the models we investigate here, all exogenous uncertainty is,
in essence, the same.  But in the actual economy, different specialists
may be regarded as different sensors and thus the exogenous events
which trigger market reactions must be treated differently.
For example, the assassination of a president probably impinges an all
and few know how to asses the information.
Information on a new court settlement concerning an insolvent firm
is picked up immediately by the bankruptcy experts and few others.
The successful testing of a new product is rarely of concern to
fundamental long term investors. 

 Modern finance tends to treat all
uncertainty as a lottery ticket.  This treatment is valuable for calculating
expectations after the odds have been assigned.  The more basic problem
is to assign the odds.  The various markets do this through experts
concerned with their interpretations of the different
qualities of risk.  As a first order approximation this can be
modelled by assuming that the experts in different areas have a greater
refinement of information in the area of their expertise than do
others.  If this were not the case, there would be no rationale for
financial institutions specializing in different forms of risk.

\section{CONCLUDING REMARKS}

Recently there has been a growth of relatively sophisticated
behavioral models in economics where the analogies to stochastic mass
particle behavior in physics and biology have been drawn.  No attempt
is made here to push analogies beyond the observation that even in the
study of equilibrium a full process model is required and that simple
rules of conservation are required.  In particular this approach
demystifies the role of money and various forms of credit which can
then be regarded as well-defined different basic particles described
by the laws of transformation.  When this procedure of modelling
economic process is adopted, the rational expectations assumption
utilized in much of micro and macroeconomics, becomes merely one among
an infinite number of behavioral procedures for updating a dynamic
process that, in general, may not be heading to a unique stationary
state.  In this model the behavioral assumptions were limited to
dividend acquisition and to simplistic behavior.  A further extension
of the type of model presented here is to have expectations of future
stock prices depend on information networking and learning, or on some
historical material and possibly on the predictions of some small
subset of experts.  

The explicit introduction of birth and death
processes for the agents provides a strong forcing function or set of
exogenous events constantly applied to the system.  A sand pile analogy
is appropriate, suppose we have a table of sand with a constant stream of
sand being poured onto it. Eventually,  if a macro
stationary state for the system as a whole is reached, the amount of
sand falling off the edges of the table
will equal that of the sand being poured
onto the pile.  Similarly the births and deaths of individuals and products may
reach balance.  But that does not tell us about the micro behavior of
the individual units.  

In our models here we have not stressed either
the overlapping generations or the expert agent aspects of an economy
with a stock market.  The rational expectations approach would have us
believe that if some agents can perceive arbitrages that others miss,
the arbitrage will quickly go away and the brighter agents will
capture all profits.  A more ecological viewpoint is consistent with
the existence of some experts who make a good living in their niches but
are not common enough and do not live long enough to capture the whole
market.  It is feasible that there are virtuoso players in finance as
there are in music. But there is little evidence that a Mozart or a
Bach can found a dynasty of progeny with equal talent for any length
of time.  The classical economic models which reflect knowing
everything and calculating everything, while learning nothing, appear
more and more to provide a poor model of stock trading.  Yet at the
other extreme a little introspection by these authors leads us to
doubt models which stress a high amount of learning.  These may err in
the other direction.  Genetic algorithms may have something to say
about evolution in the next 2,000 generations, but not to much to say
about next year's market.

Some of this work was performed while the authors were visiting the
economics program at the Santa Fe Institute.  P.B. and M.P. thank
Julie and Martin Shubik for their hospitality at their home on Cape
Cod where this work was completed.  The authors have enjoyed stimulating
discussions on economics with K. Nagel.
This work was supported  by the
U.S. Department of Energy  Division of Materials Science
under Contract No. DE-AC02-76-CH00016.

\figure{
\label{f2}
A-particles (white) diffusing from the left and B-particles (black) diffusing 
from the right.  A and B annihilate when they bump into each other near 
the center. The annihilated particles are then fed back at opposite
ends of the tube.
}
\figure{
\label{f3}
 a) Stock price variation for independent noise traders who
adjust their price randomly up and down with equal
probability (diffusion). N=500, $p_{max}$ = 
500. The time, $t$, is the average number of updates per agent. b) 
A snapshot of the
 selling prices $p_s$ and buying prices $p_b$ in the market. Note the gap 
around the current market price $p(t)$ = 247 where the number of agents is 
relatively low. c) Hurst plot of the range $R$ of price variation versus 
time, $t$. The lower curve shows the local derivative of the 
upper curve, indicating 
effective exponent $H$ of the range, $R$. It converges to  $H = 1/4$
for large $t$. The 
irregularities for large t are of statistical nature due to the 
limited data set. d) Fit to Equation (4).  The upper curve is
$\big(R^2/t^{1/2}\big)$ and the lower curve is 
$\big(R^2/(t^{1/2}\ln(t))\big)$.
The logarithm substantially flattens the small $t$ correction apparent
in the upper curve.
}
\figure{
\label{f4}
As figure 2, but with an additional drift D=0.05 towards the market price.
}
\figure{
\label{f6}
The urn model.  Data shown as in
 Figure 2. N=200, D=0.05. The agents copy existing 
traders when selecting new asking prices and bidding prices. The price 
fluctuations are much more dramatic here, and the resulting 
asymptotic exponent for price variations is equal to the random walk
value of $H= 1/2$ rather than  $H= 1/4$.  Note that the width of the 
distribution of prices in the market (Fig. 4b) is now self-organized,
rather than limited by the boundary conditions on $p$.
}
\figure{
\label{f7}
Volatility feedback in the urn model. The diffusion of prices for each
agent is equal to the observed price variation over the previous 50
time steps. N=1000, D=0.2. Note the wide plateau where the exponent
$H=0.65$. d) Distribution of price variations for various time intervals
$dt$. e) Scaling plot of price fluctuations as defined in text. All
curves can be described by a single scaling variable. The scaling
exponent is $H=0.65$. f) Price fluctuations for a single value of
$dt=1600$ compared with a Gaussian.  Note the fat tail, indicating a
significant probability of having variations exceeding several
standard deviations.  }

\figure{
\label{f8}
Urn model with 2\% of the agents are fundamental value buyers,
and N=500. 
 The fundamental value traders ask and bid prices in the range
between 1931 and 2068. a) Note the ``bubble'' around t=200000 where the market
price exceeds anything that can be justified by fundamental value 
estimates. b) A typical configuration of prices in the market.
}
\figure{
\label{f9}
a) Price variations for 20\% fundamental value traders. Market prices are
 confined to the region
 spanned by the fundamental value traders, except for short glitches. 
 b) Distribution of prices at the anomalous low price around 
t=400000. At that point, the distribution is skew, with none of the rational 
traders owning shares.
}

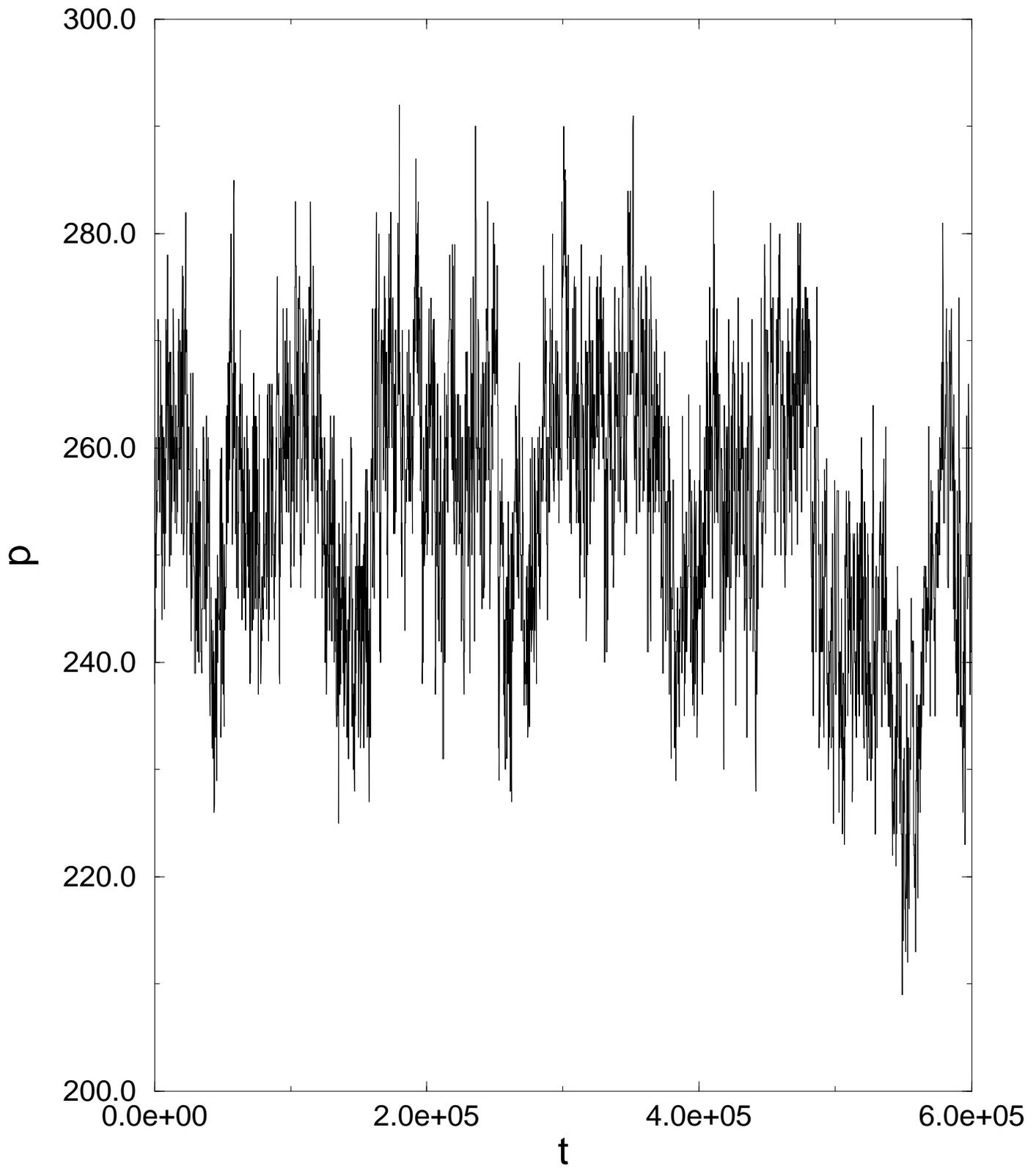
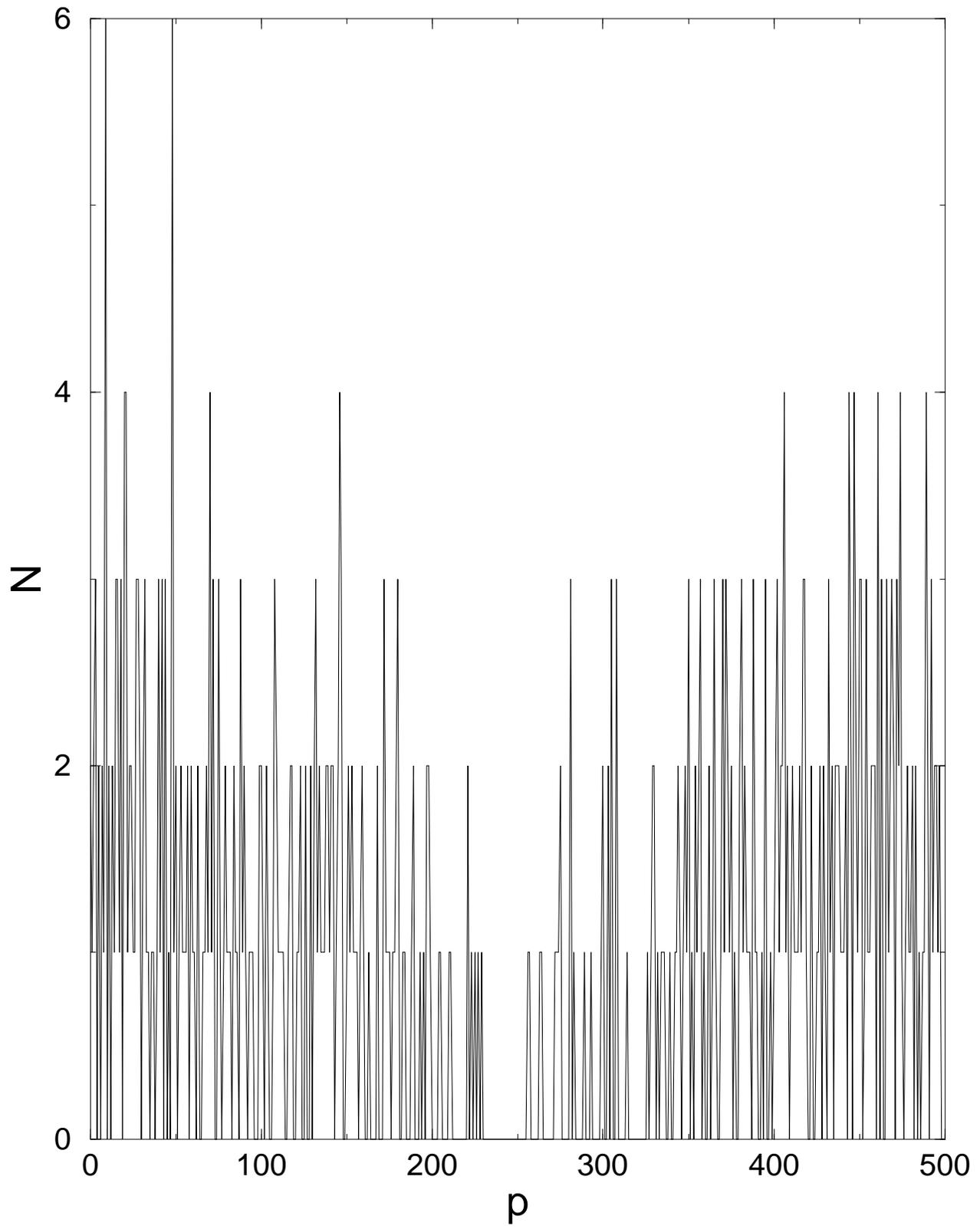
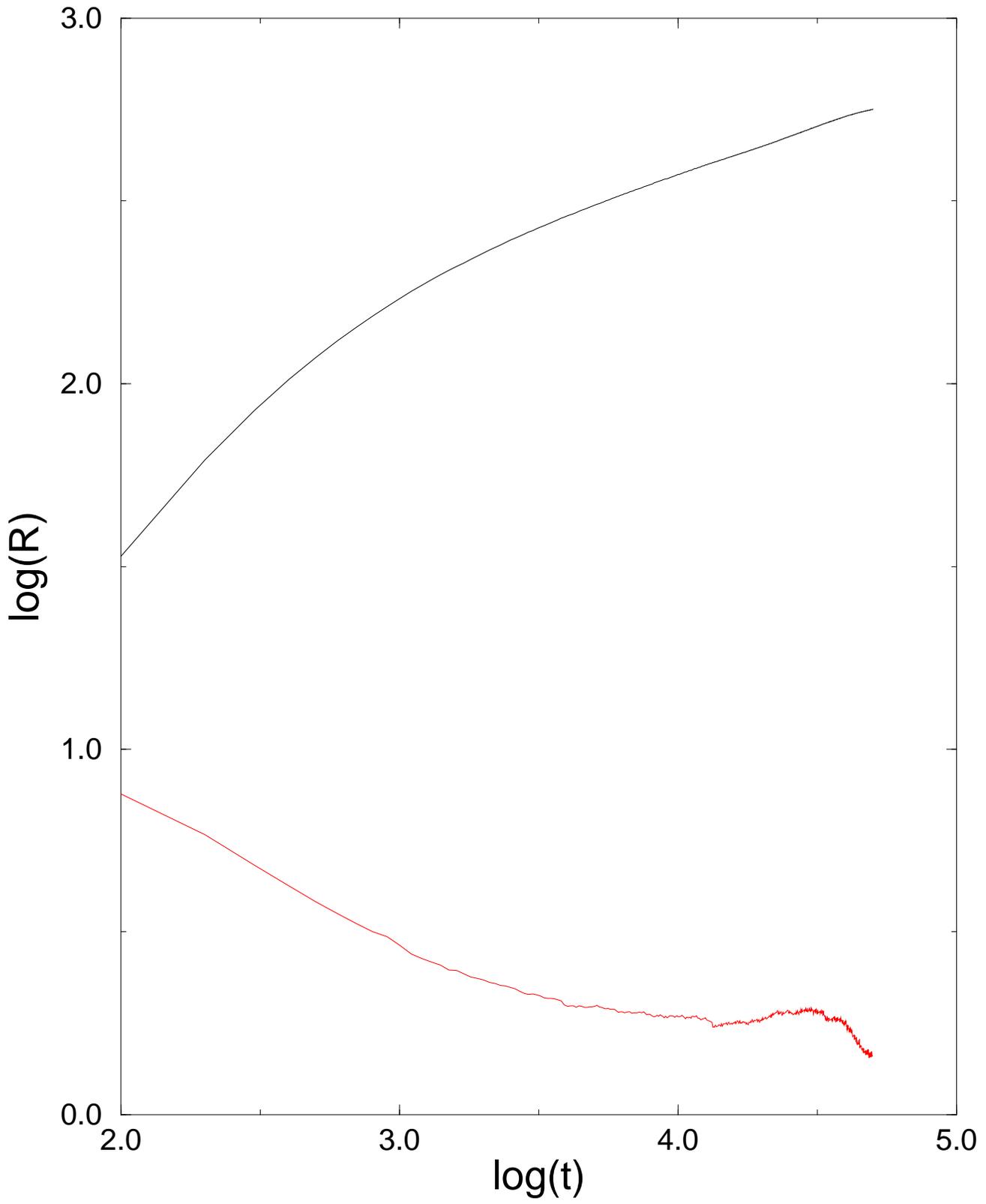
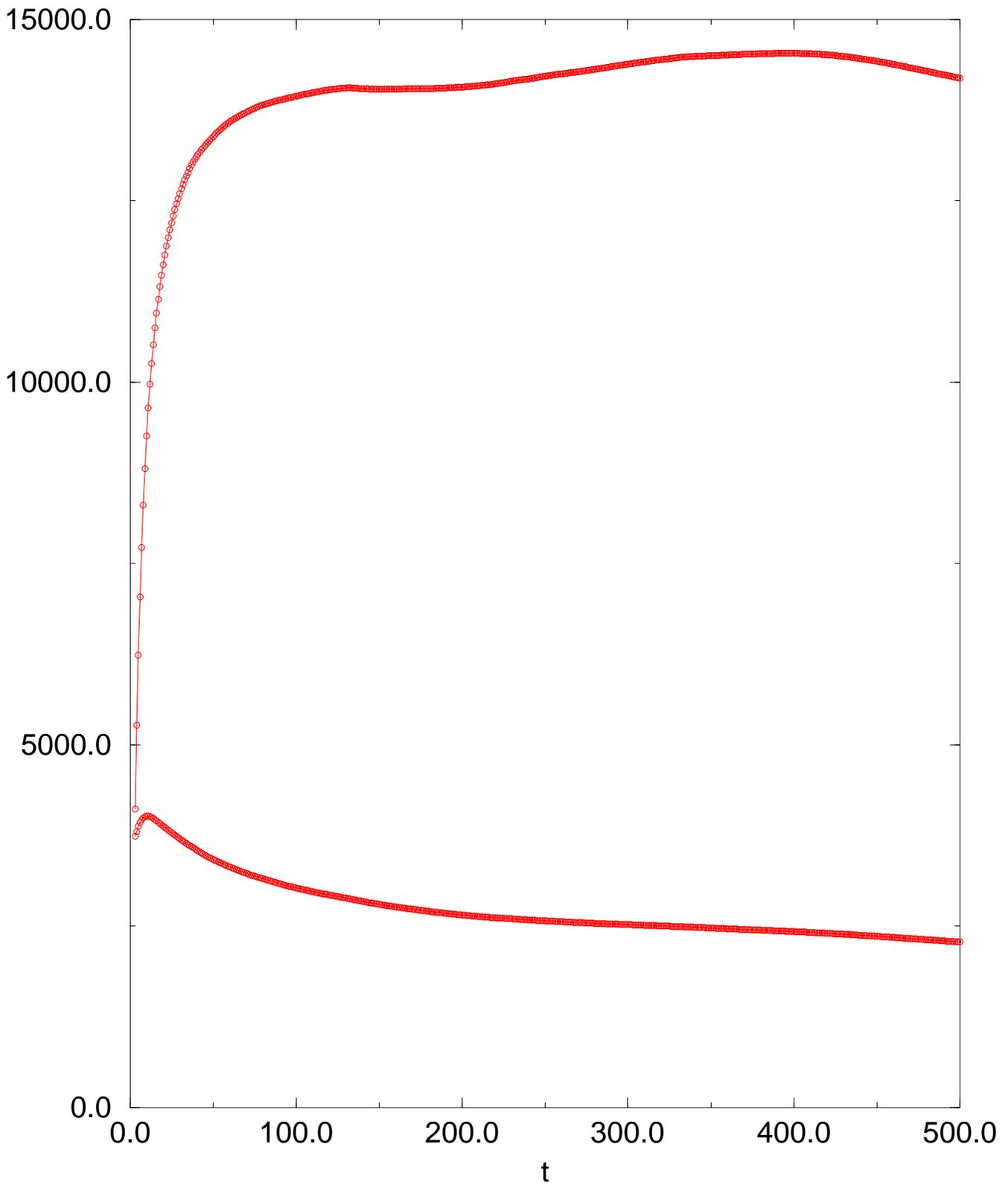
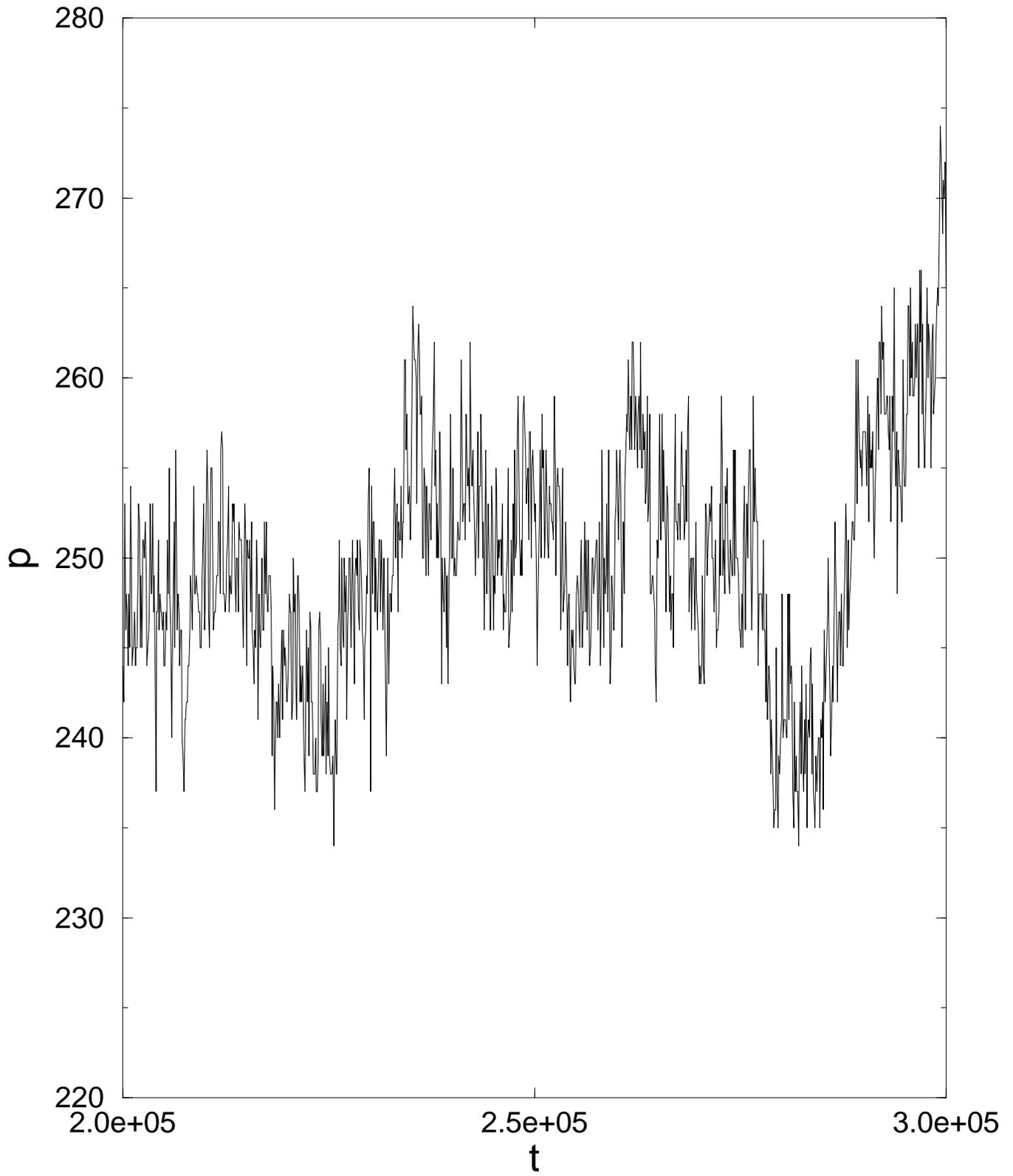
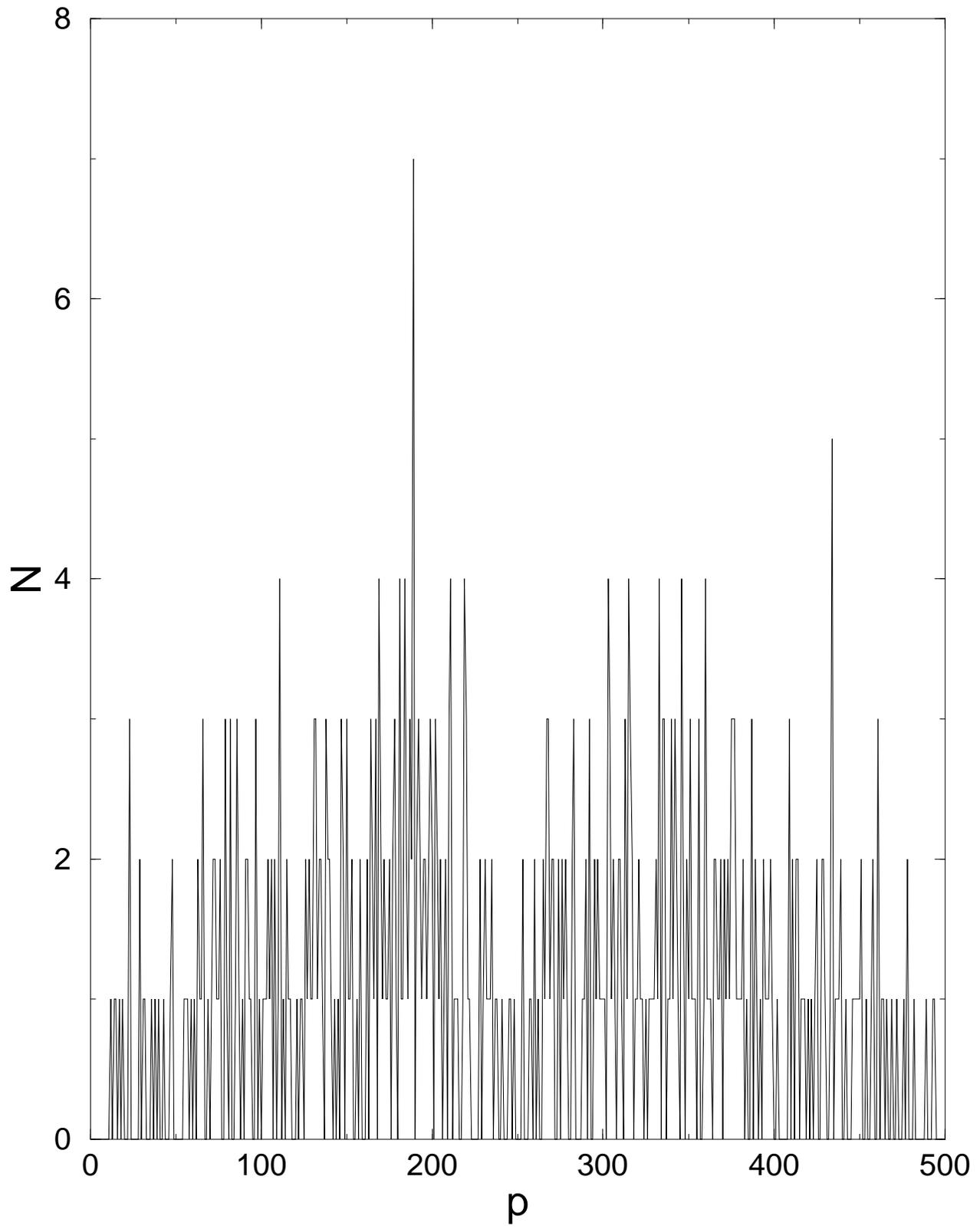
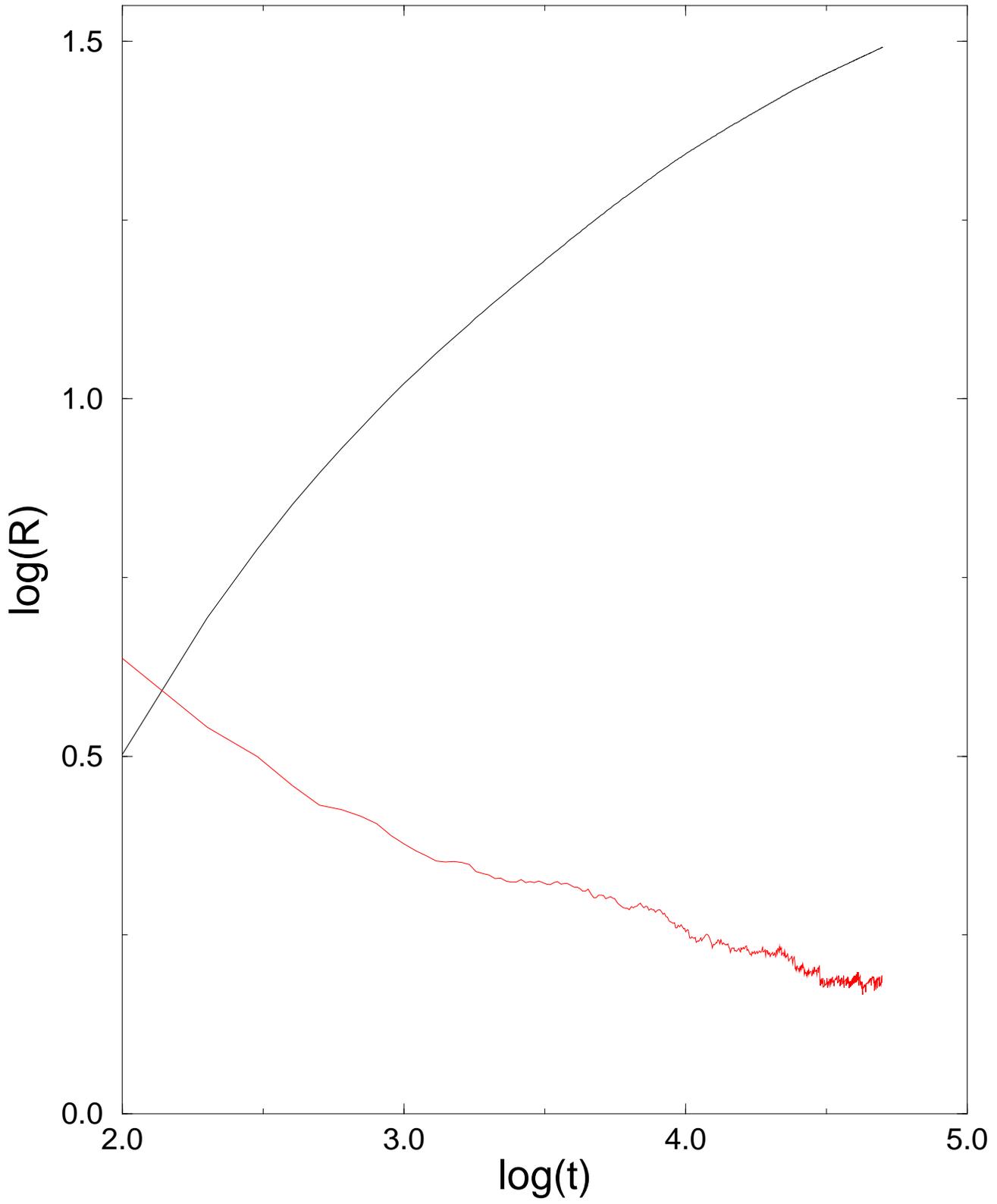
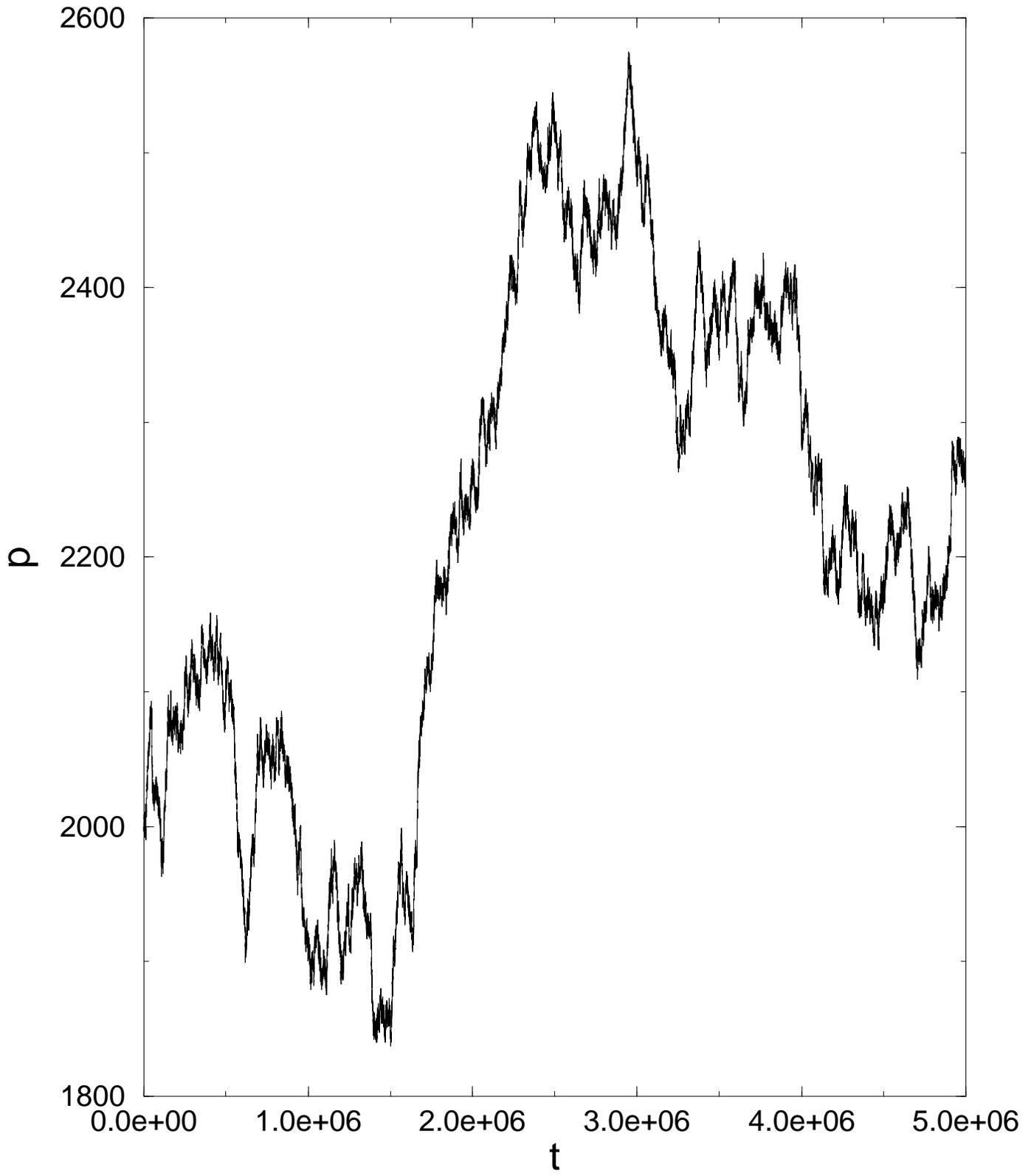
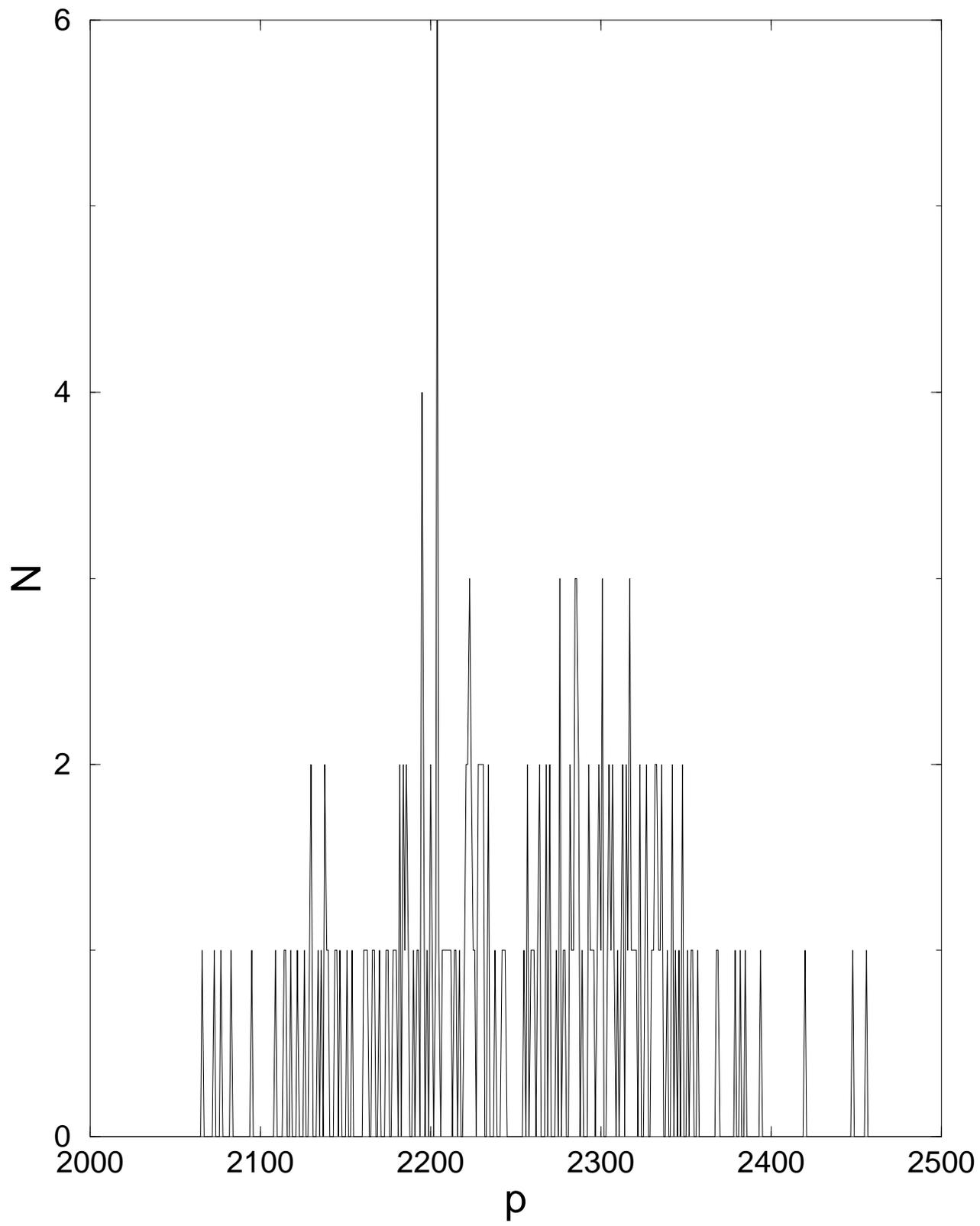
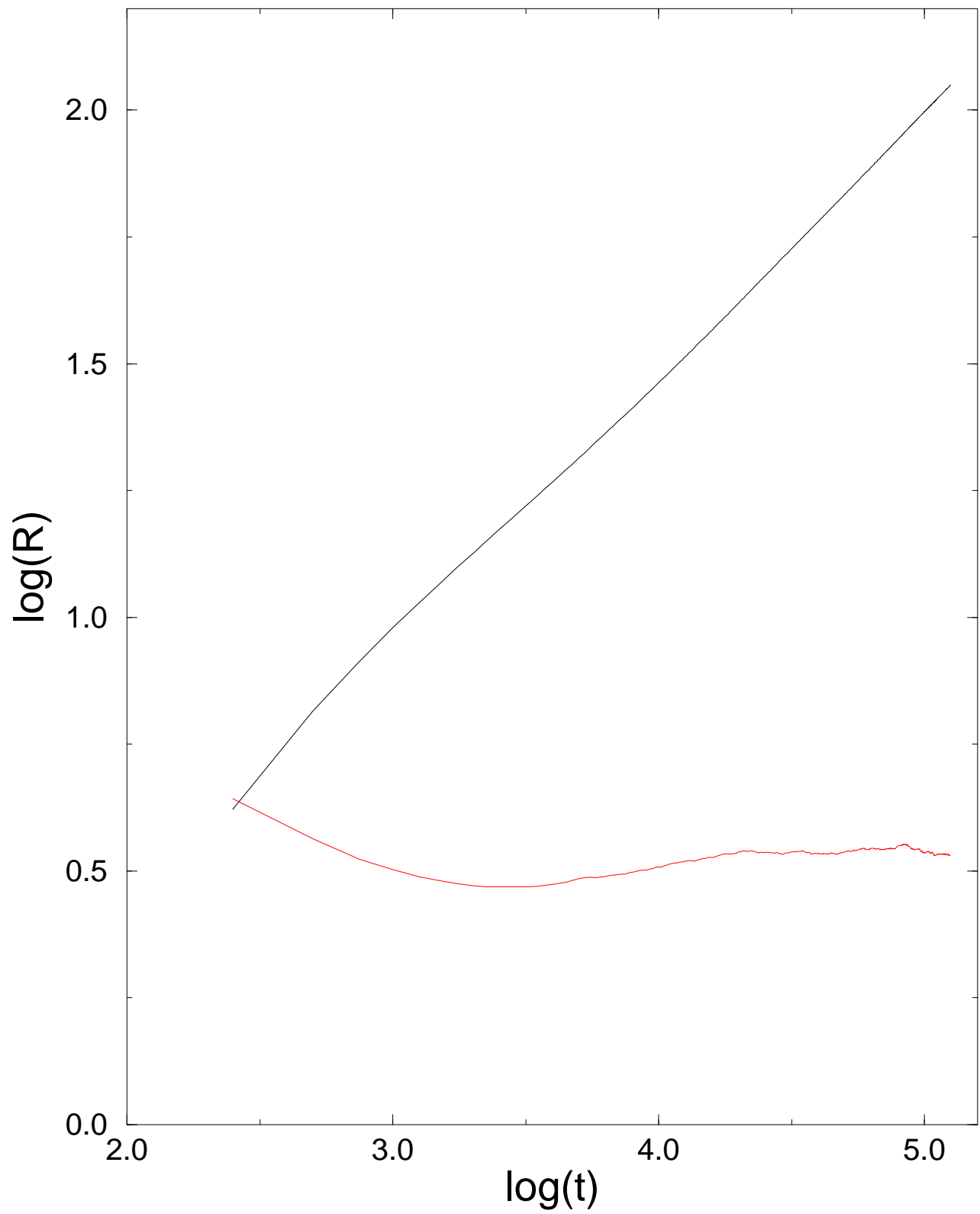
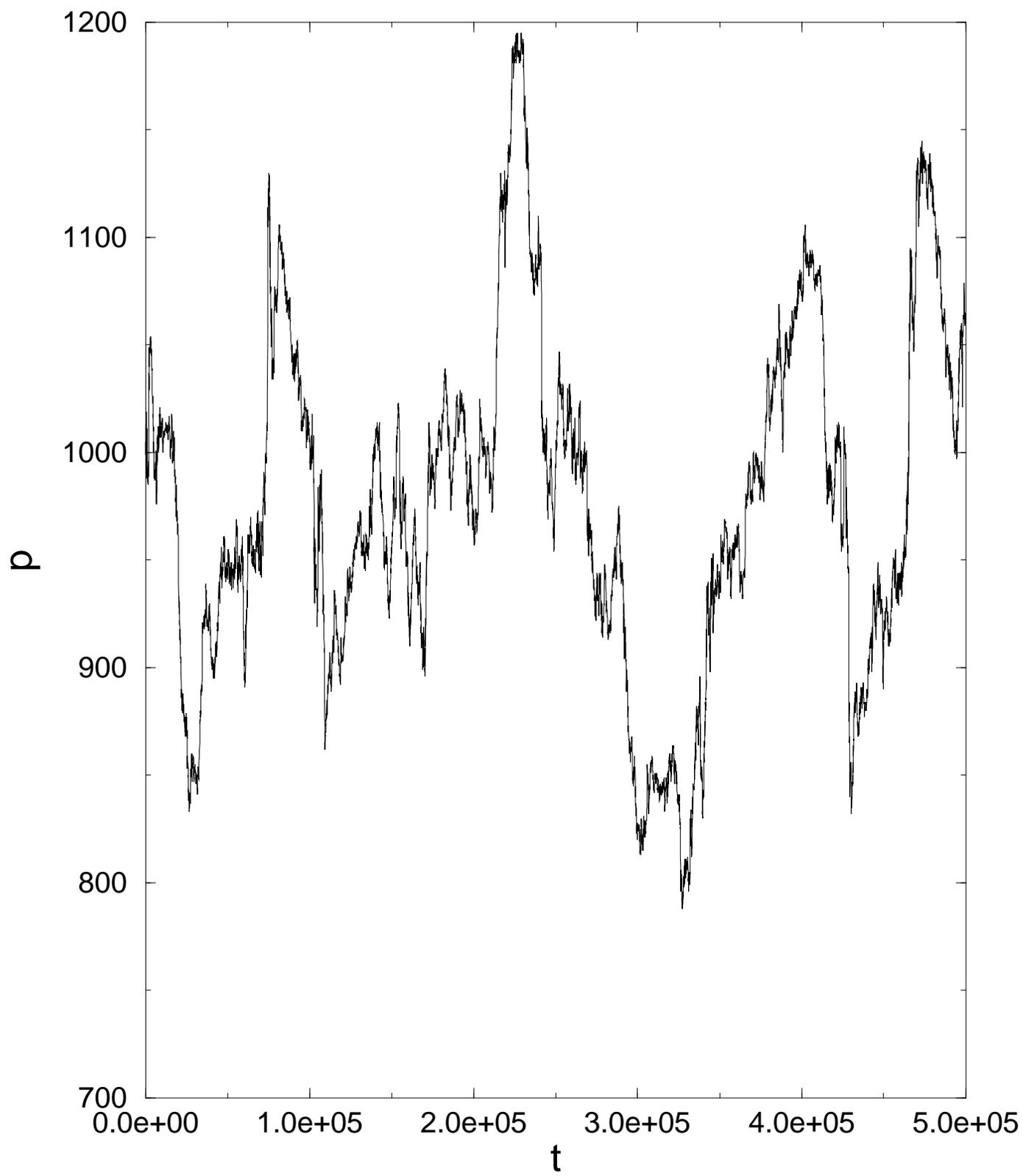
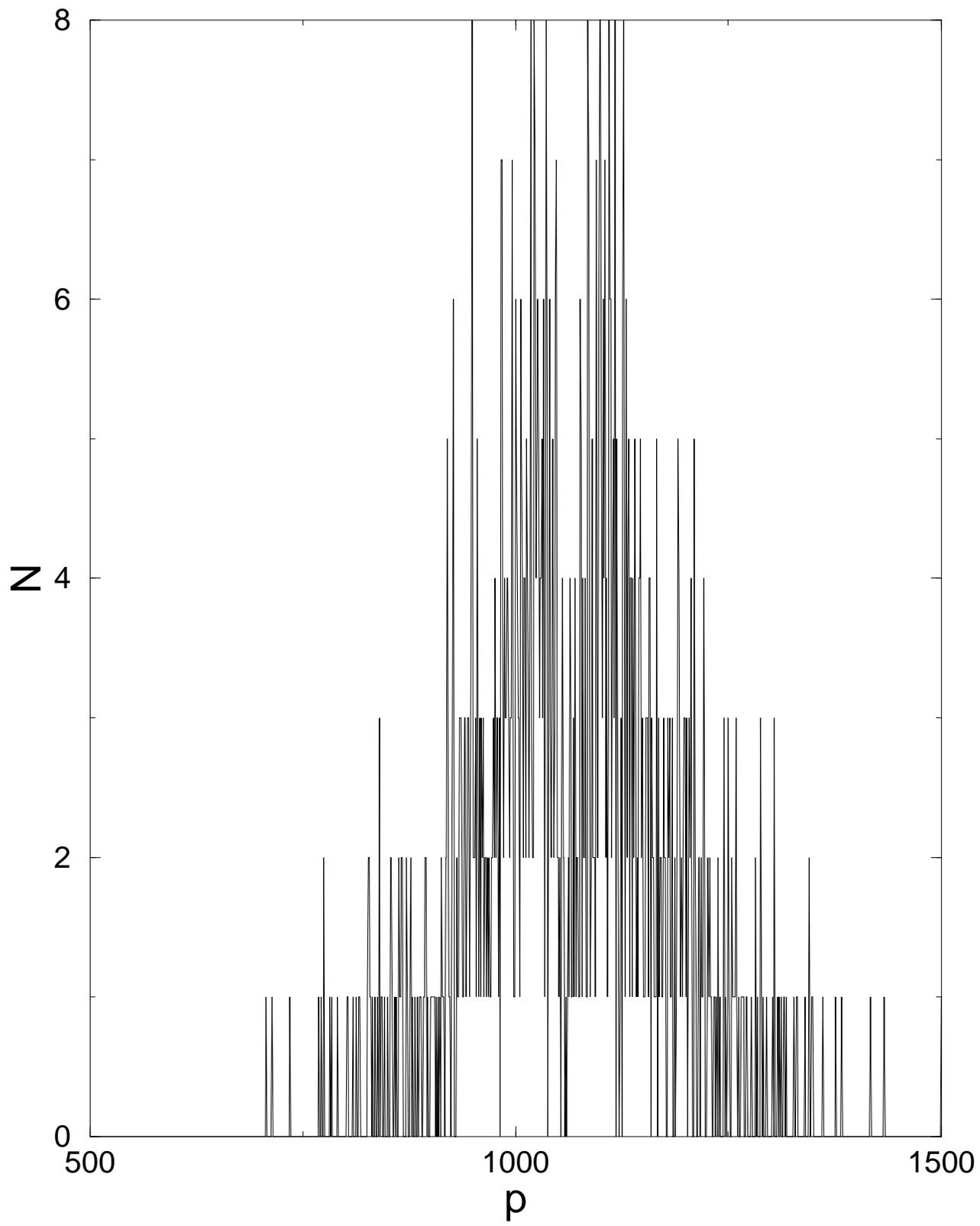
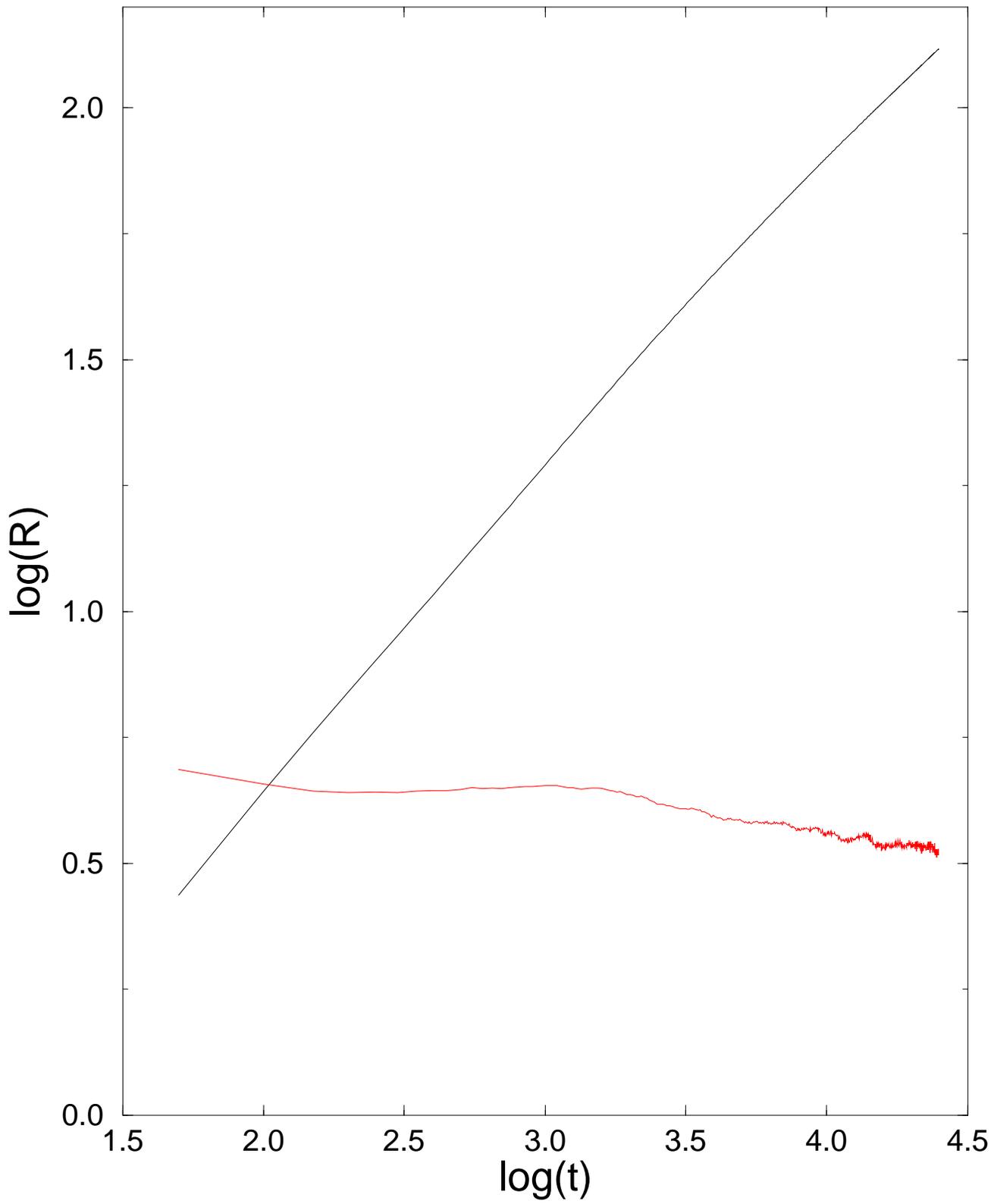
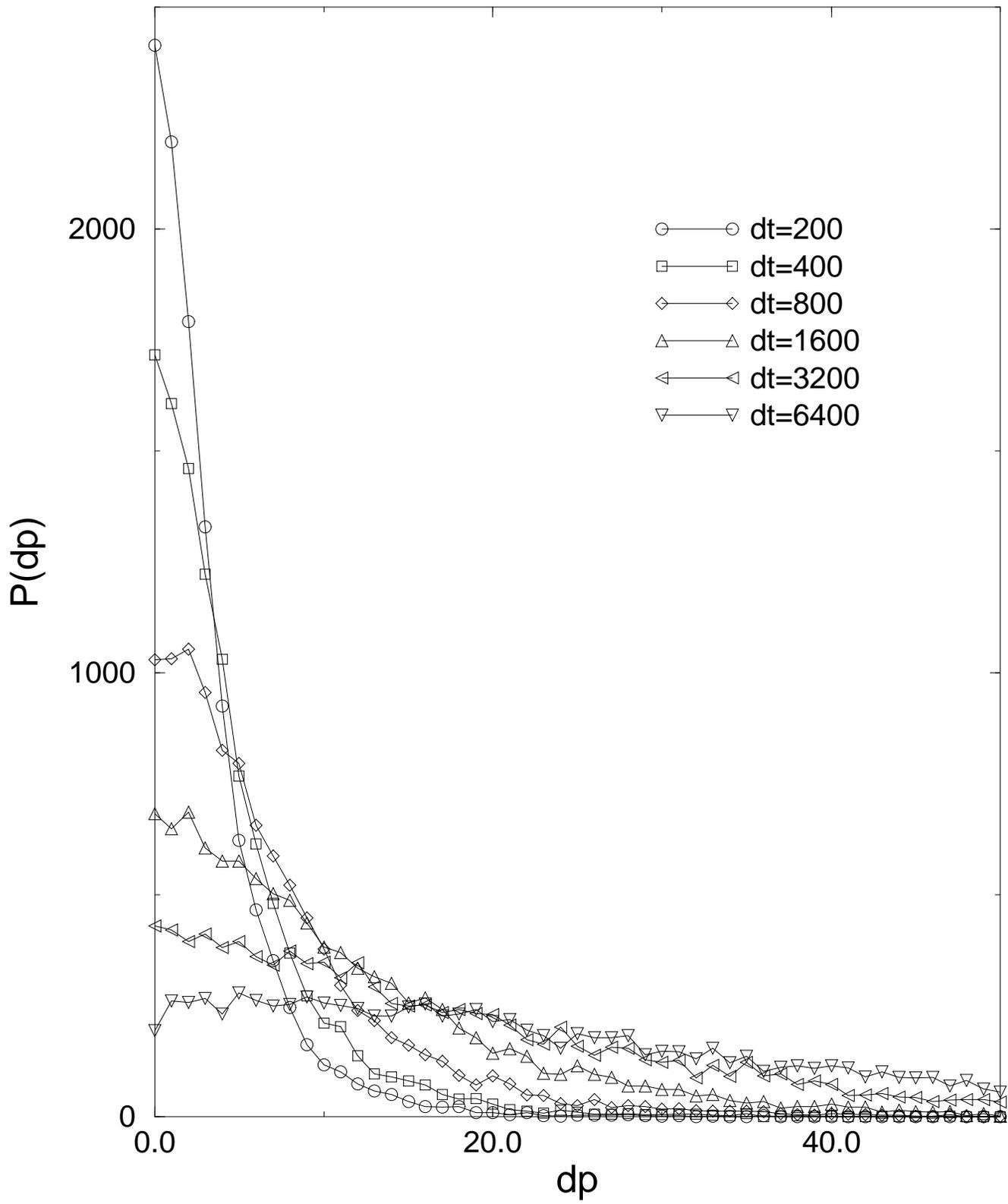
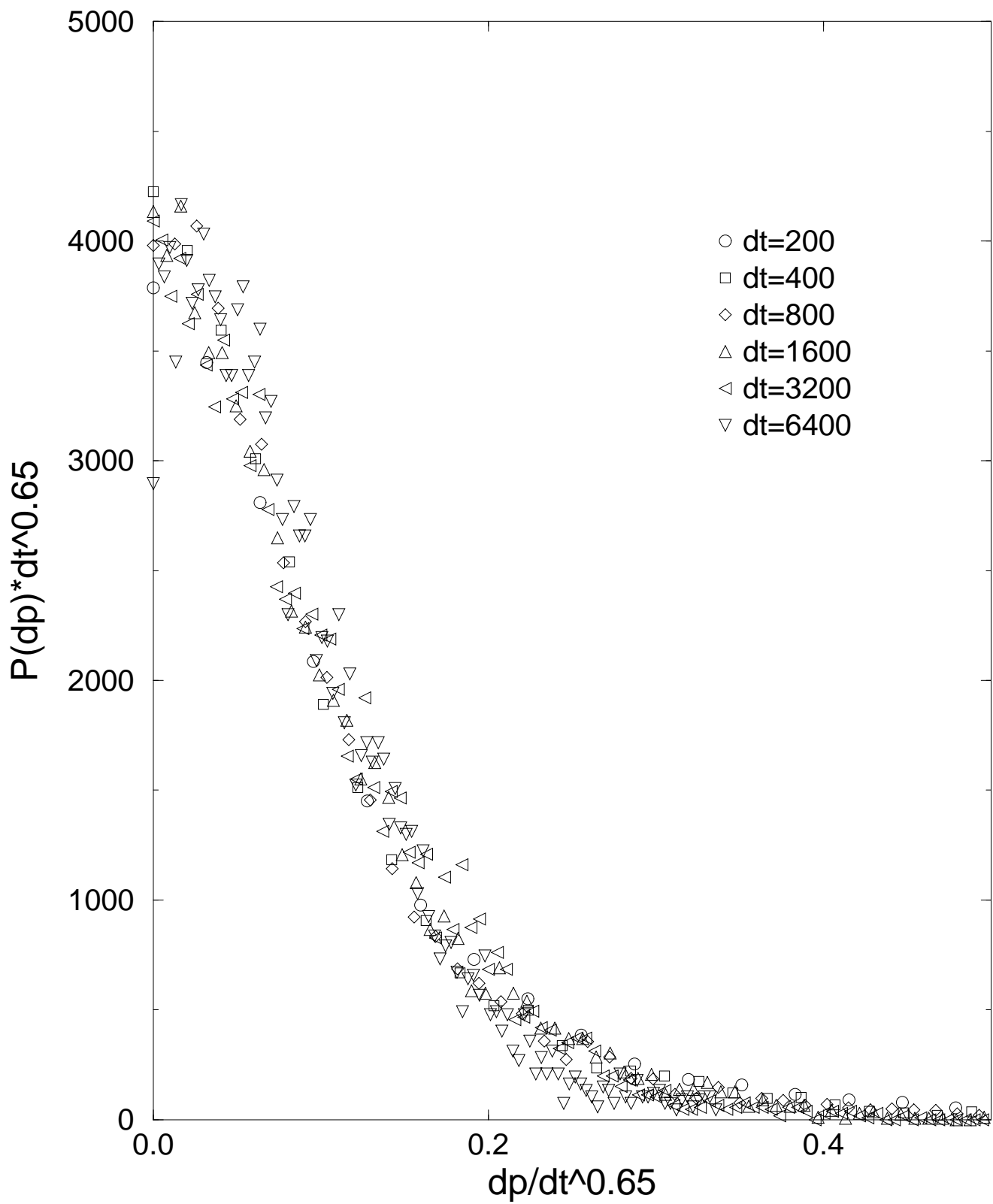
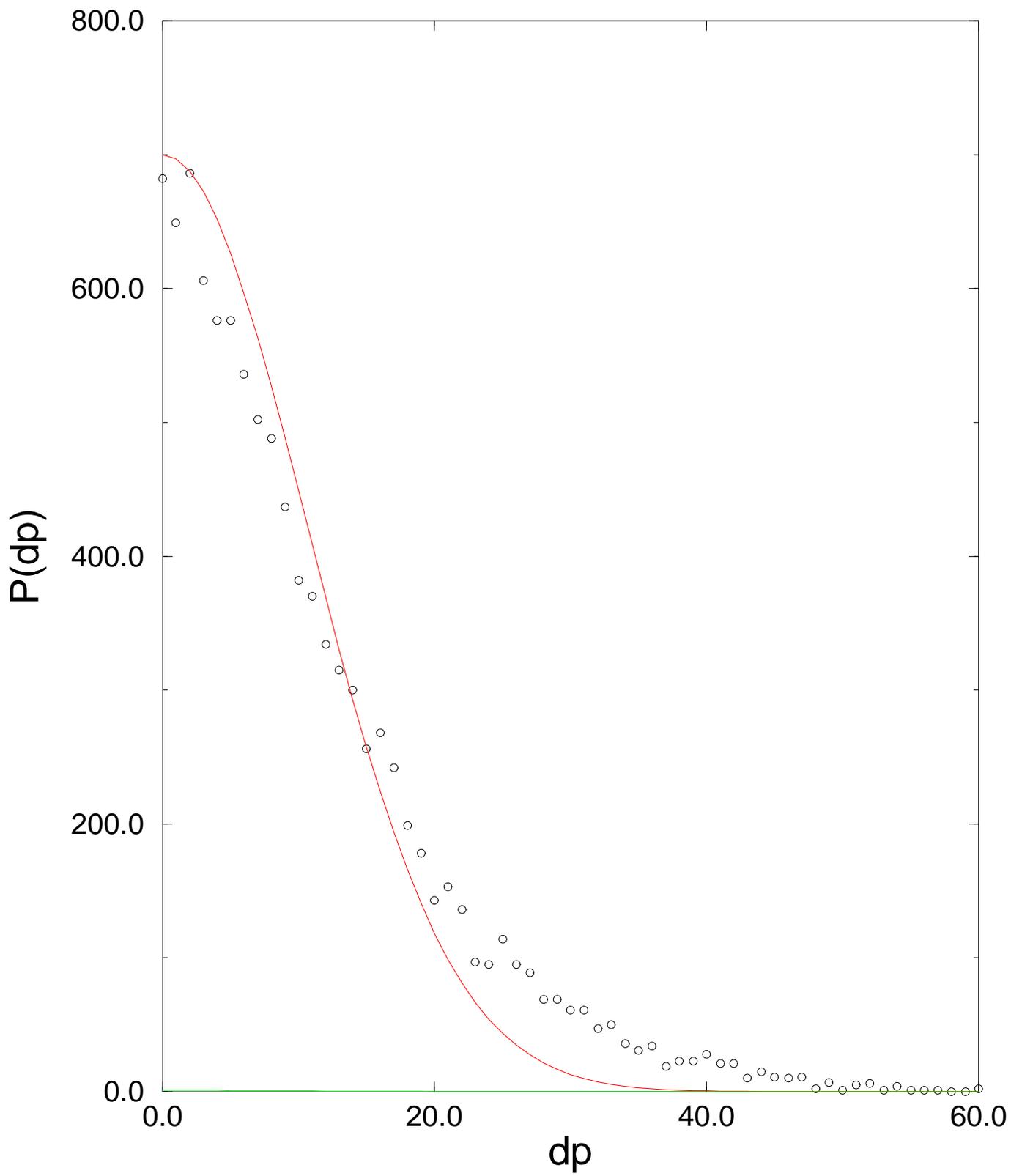
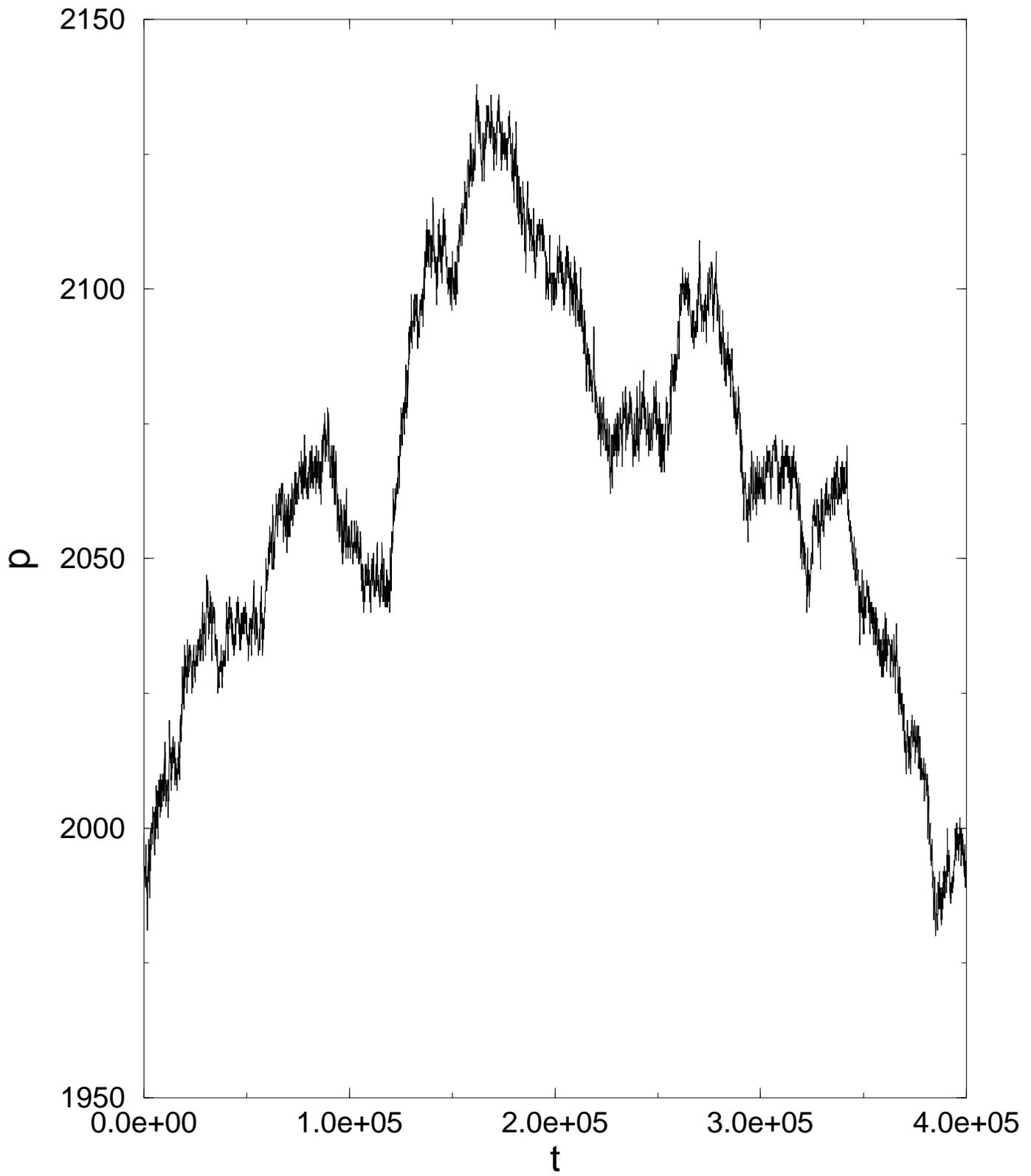
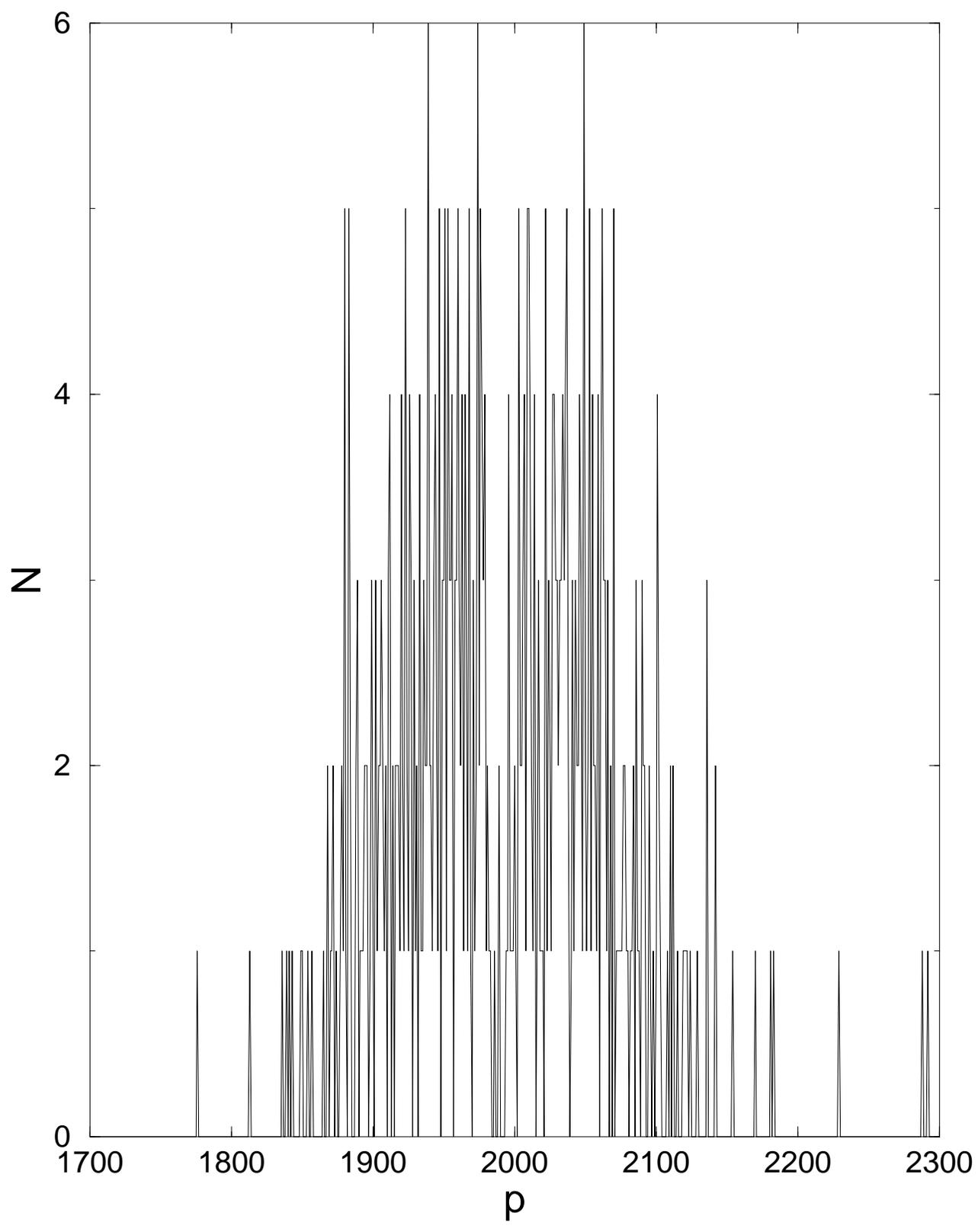
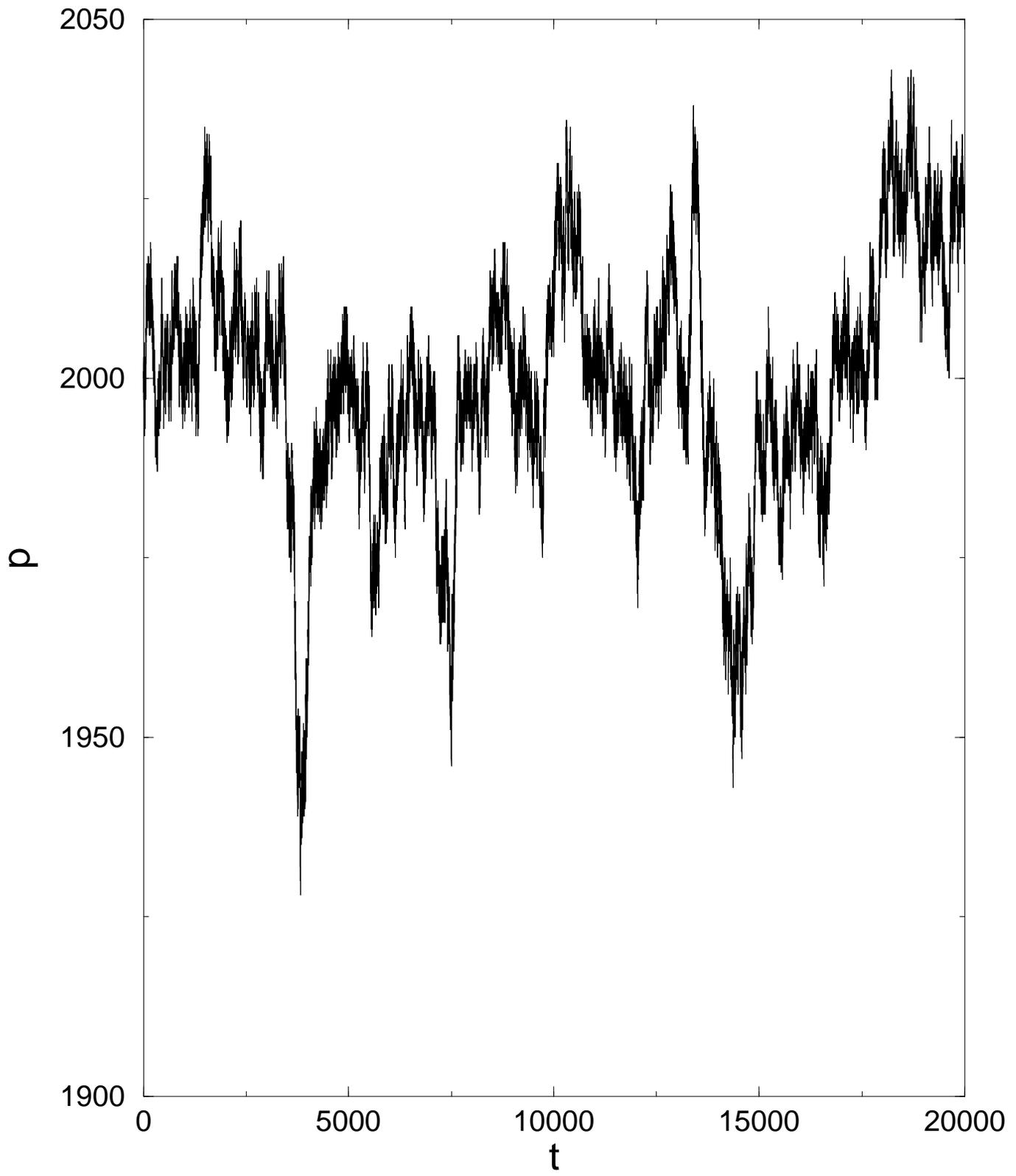
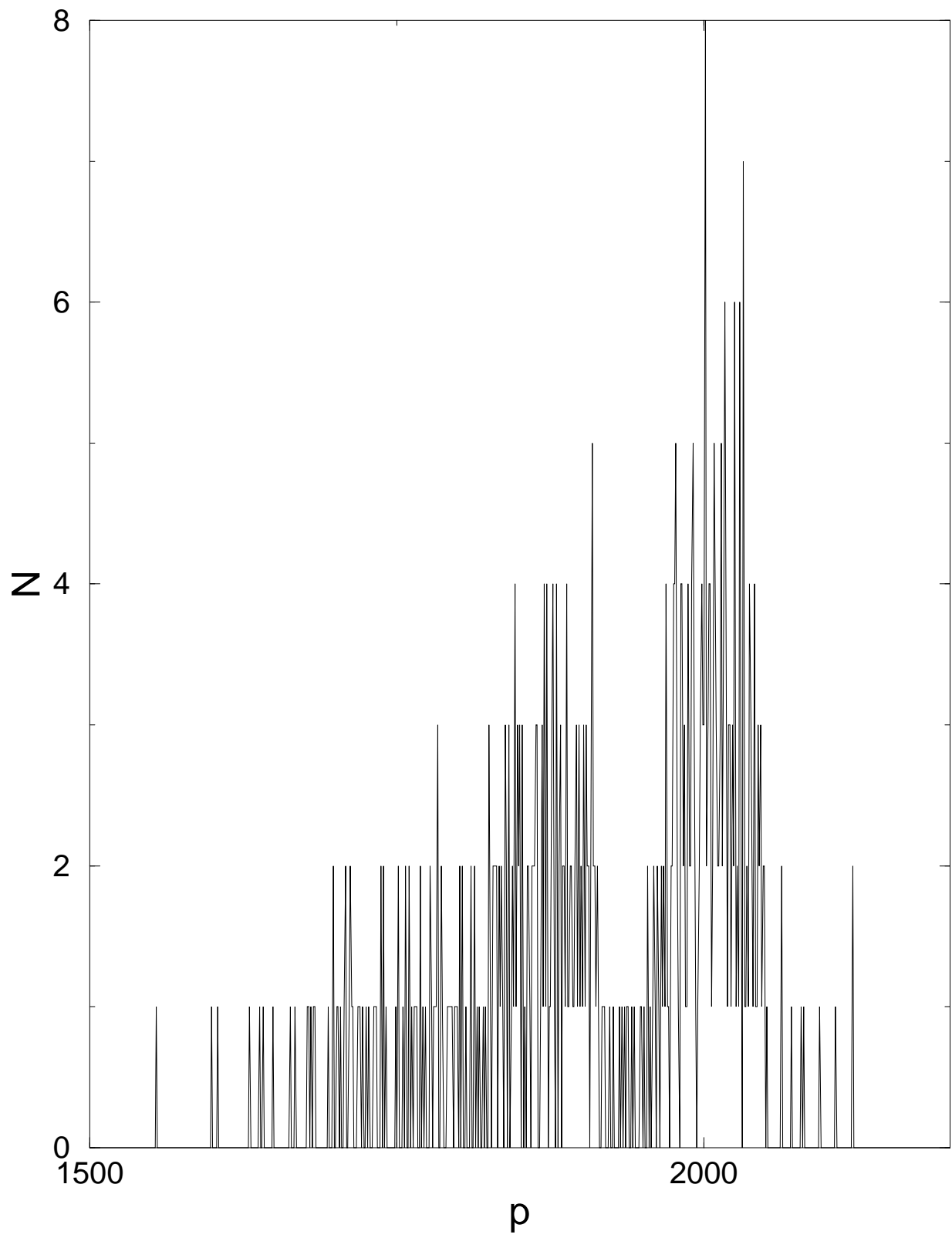
\end{document}